\documentclass[twocolumn]{article}
\title{Modelling Dynamic Limitations of Lithium Transport in Lithium-Sulfur Batteries}
\author{Dr. Michael Cornish, Dr. Monica Marinescu, Mr. Dharshannan Sugunan}

\usepackage{graphicx} 
\usepackage{amssymb} 
\usepackage{cite} 

\usepackage{color} 

\begin{document}
\maketitle

\abstract{A common, yet unmodelled experimental phenomenon is explained first through the experimental data and then through modelling efforts. The phenomenon is a large, temporary loss in voltage during constant current discharge. This effect can be observed in several independent experimental settings, such as during low-temperature operation, under high currents, with low Electrolyte-to-Sulfur (E/S) ratios, as well as in cells with high salt concentrations. The model presented here is the first to prove to capture this effect. Moreover, several other related experimental results are presented, discussed, and predicted by the model. A fundamental explanation for why most Li-S models are unlikely to be able to capture the large voltage loss effect is also given.
}
\section{Introduction}

Among the top candidates to enhance the energy storage sector, Lithium-Sulfur (Li-S) batteries have been steadily gaining in research prominence. Li-S batteries stand as a promising beyond Li-ion battery technology due to a theoretical specific energy of 2567 Wh/kg.  Compared to Li-ion batteries, this is an order of magnitude improvement \cite{Bruce2012}. Additionally, the comparative environmental damage of Li-S cell production is expected to be far lower \cite{Song2013}. However, Li-S batteries have proven to contain complex internal dynamics that are difficult to quantify with certainty. A theoretical specification of the underlying mechanisms that most impact cell performance would greatly enhance efforts to improve the cell design and operation. Therefore, research into physics-based mathematical models of Li-S cells have coincided with the experimental investigations.

Specifying the complex Li-S cells through mathematical models allows researchers to precisely specify their hypotheses regarding the underlying mechanisms. Therefore, more precise predictions of these hypotheses are enabled, allowing for the research community to have greater certainty in the mechanistic understanding of this enigmatic storage system. Once a primary set of mechanisms is shown to accurately predict cell behaviour, work on optimisation of cell operation can be made. Moreover, dependent upon the trust in the model formulation, optimisation of cell design can also be informed through numerical investigations. However, before the benefit of modelling can be reaped, the modelling community must contend with and reproduce at least the most common experimental results found throughout the literature. 

In this work, Section \ref{sec:Current-Temperature Isometry} outlines what will be referred to as the Current-Temperature Isometry. This is an extension of an experimental phenomena observed throughout the literature. Section \ref{sec:Dynamic Limitations} lays out experimental evidence for the proposed model of this work along with the argument for why all models in the literature are unable to capture the Current-Temperature Isometry. In particular, Section \ref{sec:Precipitation Mechanisms} explains why precipitation and its consequences are fundamentally insufficient. Section \ref{sec:Transport Limitations} details the transport limitation models in the literature along with the required adjustments to these mechanisms and their associated electrochemistry. Section \ref{sec:Model Development} outlines the proposed mathematical model. Section \ref{sec:Model Validation} discusses the validation procedures used in the parameterisation of this model, as recommended in Cornish \& Marinescu \cite{cornish2022toward}. Section \ref{sec:Results & Discussion} discusses the results of the model. Concluding remarks regarding the implications of this work are given in section \ref{sec:Conclusion}. Further details to supplement the arguments within the main body of this work are given in Section \ref{sec:Supplementary Material}. 

\section{Current-Temperature Isometry}\label{sec:Current-Temperature Isometry}

In the experimental work of Hunt et al. \cite{Hunt2018}, single layer pouch cells were cycled with a constant current under a constant temperature. In particular, the cells were held at either 20$^{\circ}$C, 30$^{\circ}$C, or 40$^{\circ}$C through the use of a Peltier element. All three temperatures yielded Ohmic resistance curves with a peak during the discharge plateau transition. Generally speaking, Ohmic resistance increased with decreasing temperature. While these discharge Ohmic resistance curves were qualitatively similar, their associated voltage profiles differed greatly. 

\begin{figure}[h]
\includegraphics[width=\columnwidth]{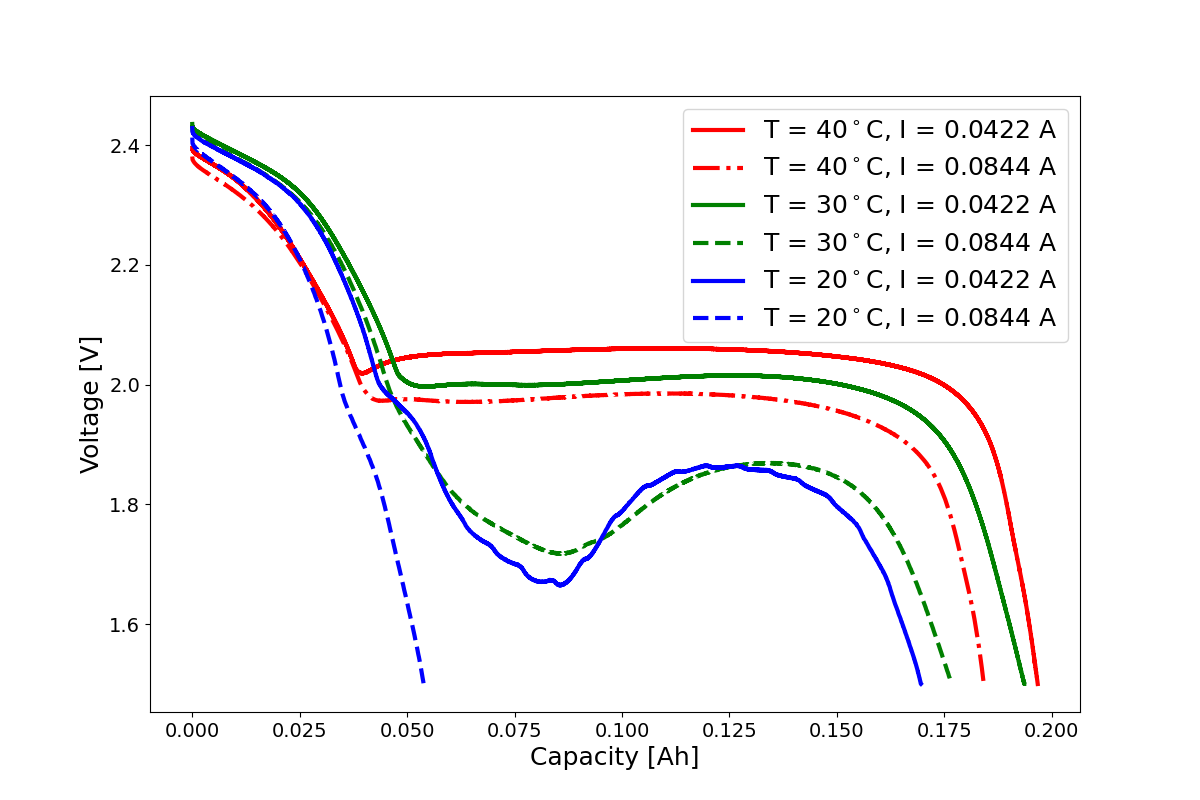}
\caption{Hunt et al. \cite{Hunt2018} experimental data. The originally published data only provided the 0.2C discharge voltage curves while this figure includes the 0.4C discharge curves at all temperatures. The 30$^\circ$C, 0.4C and 20$^\circ$C, 0.2C clearly display the Current-Temperature Isometry.}
\label{Fig:HuntData-ALL}
\end{figure}

Originally published in Hunt et al. \cite{Hunt2018}, the 0.2C discharge of the 20$^{\circ}$C, reproduced in Figure \ref{Fig:HuntData-ALL}, clearly yields an unusual voltage profile. However, unpublished in the original article was the 0.4C discharge of the 30$^{\circ}$C cell. As shown in Figure \ref{Fig:HuntData-ALL}, these two voltage curves overlap remarkably well. Rather than an experimental fluke, this work aims to explain that the similarities reflect an underlying limiting mechanism.

\begin{figure}[h]
\includegraphics[width=\columnwidth]{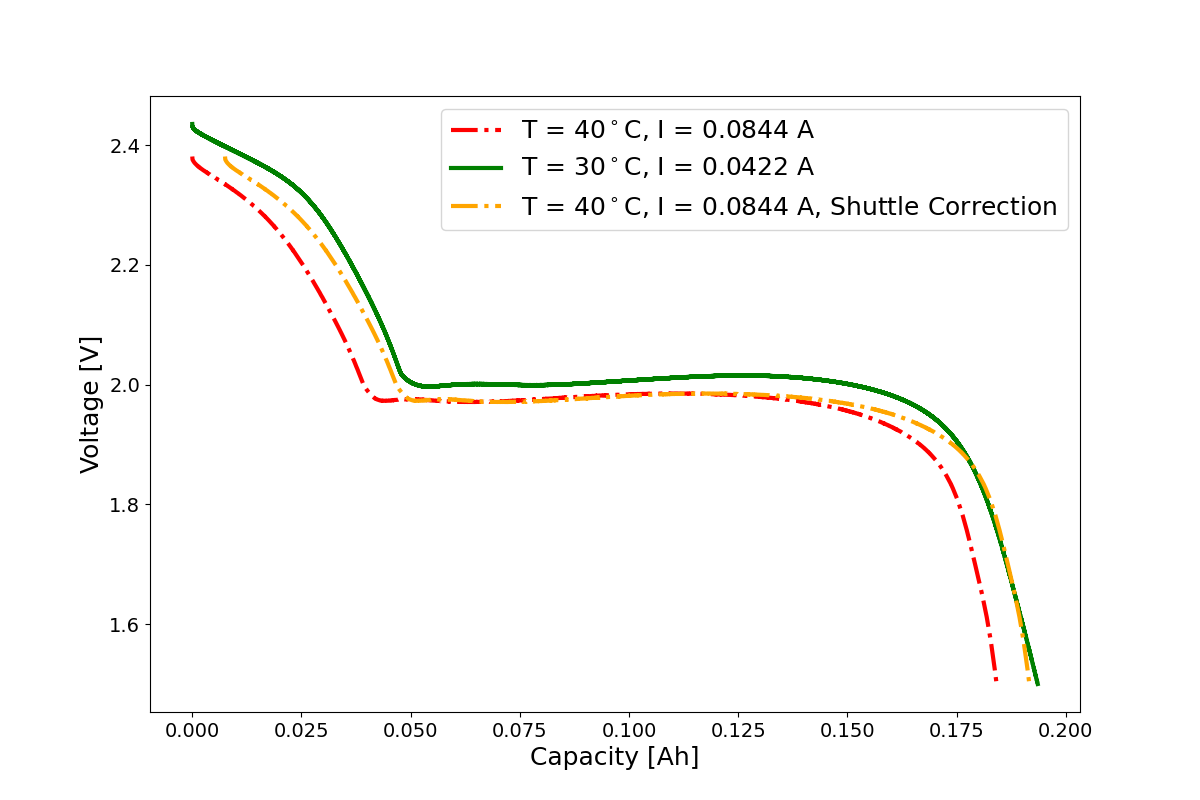}
\caption{Shuttle correction to the 40$^\circ$C, 0.4C discharge voltage curve from Hunt et al. \cite{Hunt2018} experimental data. By adjusting for the different state of charge due to shuttle, the Current-Temperature Isometry can be more easily observed. A slightly lower current than 0.4C is likely to yield a more precise match with the 30$^\circ$C, 0.2C voltage curve. }
\label{HuntData-ShuttleAdjusted}
\end{figure}

Similarly, there is a connection between the 40$^{\circ}$C cell discharged at 0.4C and the 30$^{\circ}$C cell discharged at 0.2C. However, this data is complicated by several factors. First, due to the higher temperatures, more shuttle is observed and therefore the state-of-charge at the outset of discharge creates a misleading voltage profile. The 0.4C, 40$^{\circ}$C discharge profile is adjusted to reflect the state-of-charge loss due to shuttle in Figure \ref{HuntData-ShuttleAdjusted}. The adjustment amounted to aligning the final capacities. Second, the 0.2C, 30$^{\circ}$C low plateau voltage is consistently higher than that of the 0.4C, 40$^{\circ}$C. Although the affects of doubling the current and increasing the temperature by 10$^{\circ}$C seemed to balance out in the case of the 0.2C, 20$^{\circ}$C and 0.4C, 30$^{\circ}$C case, the same relationship is not exact in the 0.4C, 40$^{\circ}$C and 0.2C, 30$^{\circ}$C case. From the 0.2C, 40$^{\circ}$C voltage profile in Figure \ref{Fig:HuntData-ALL}, it is clear that increasing the current does not alter the high plateau voltage greatly but does alter the low plateau voltage significantly. Therefore, implementing a slightly lower current than 0.4C and utilising the same procedure for state-of-charge adjustment due to shuttle is likely to yield a more precise match between the 0.4C, 40$^{\circ}$C voltage curve to that of the 0.2C, 30$^{\circ}$C. 

There is no data available to compare a higher discharge current for the 30$^{\circ}$C and 40$^{\circ}$C cells. However, it is reasonable to extrapolate from the above analysis to state that at approximately 0.8C the 30$^{\circ}$C cell may match the 0.4C, 20$^{\circ}$C voltage curve. If this is the case, then because the 0.4C, 40$^{\circ}$C cell behaves as the 0.2C, 30$^{\circ}$C cell, a 1.6C, 40$^{\circ}$C voltage curve will also match the 0.8C, 30$^{\circ}$C and thus will also match the 0.4C, 20$^{\circ}$C voltage curve. Lowering the current of the lower temperature cells to match the higher temperature cells at relatively higher currents is likely to cause a similar relationship. Roughly, if the temperature of the cell is changed by adding $\delta \times 10^{\circ}$C, $\delta\in\mathbb{R}$, then multiplying the discharge current by $2^\delta$ will yield similar results after adjustment for shuttle. It is this trade-off that will be referred to as the Current-Temperature Isometry.

\section{Dynamic Limitations}\label{sec:Dynamic Limitations}
Regardless of the validity of the Current-Temperature Isometry, a pattern extrapolated from limited data, the voltage profiles shown in Figure \ref{Fig:HuntData-ALL} have yet to be explained. Much of the modelling literature focuses on precipitation dynamics to explain both voltage and capacity losses during discharge. However, the data in question cannot be fully explained by the secondary effects of precipitation such as pore blocking or active surface area covering. The faster such dynamics are, the less pronounced the dip-and-recovery feature will be. 

\subsection{Precipitation Mechanisms}\label{sec:Precipitation Mechanisms}

As shown in Marinescu et al. \cite{Marinescu2016}, faster precipitation dynamics should lead to higher low plateau voltages. In their work, a slow nucleation model is employed that is widely used throughout the modelling literature. With no precipitation, the lower plateau drops greatly. With increasing precipitation there can occur a voltage dip and then a recovery to a higher low plateau voltage. The precipitation dynamics remove the end-products of the electrochemical reactions, ensuring the low plateau electrochemical reactions remain relatively favourable; hence the higher lower plateau. The intermediate dip between high and low plateaus is due to the slow nucleation, which initially inhibits the precipitation dynamics during the start of the low plateau dynamics. Once nucleation limitations are overcome, precipitation clears end-products of the electrochemical reaction causing a higher voltage.  

Such a dip occurs for the 0.2C, 40$^{\circ}$C discharge voltage shown in Figure \ref{Fig:HuntData-ALL}. Lower temperatures should increase the rate of particle growth. Therefore, the lower temperature cells should be less limited by precipitation effects. Indeed, there is no dip and recovery for the 0.2C, 30$^{\circ}$C cell. Lower voltages should increase the rate of nucleation \cite{fan2015mechanism}. The lack of a dip and recovery during the 0.4C, 40$^{\circ}$C is likely due to a balancing of heightened nucleation, caused by some transport limitations lowering the voltage, and limited particle growth, caused by the higher temperature. The dip and recovery for the 0.4C, 30$^{\circ}$C cell and the 0.2C, 20$^{\circ}$C are therefore unlikely to emerge from the first order effects of precipitation because the particle growth rate should increase with lower temperatures and the nucleation rate should increase with lower voltages. However, interpretation of this experimental data through   zero-dimensional models as found in \cite{Marinescu2016}, \cite{Marinescu2018}, \cite{Hua2019}, \cite{cornish2022toward} is constrained by the lack of modelled second order effects such as pore blocking and active surface area loss. 

The more sophisticated model from Kumaresan et al. \cite{Kumaresan2008} also yields similar conclusions. In this and similar models, precipitation leads to pore blocking and active surface area loss. As shown in Ghaznavi \& Chen \cite{Ghaznavi2014b}, Figure 3 (d,e), increasing the precipitation rate of either $Li_2S_2$ or $Li_2S$ results in a narrowing and flattening of the dip-and-recovery feature. Within their work, Figure 3 (f) shows that given a fixed reaction rate, increasing the current generally flattens the low plateau. This aligns with the Hunt data  in Figure \ref{Fig:HuntData-ALL} for the 40$^{\circ}$C cell at 0.2C and 0.4C but not the 30$^{\circ}$C and 20$^{\circ}$C cells. Moreover, increasing the initial porosity doesn't alter the qualitative features of the discharge curve across currents (Figure 6, \cite{Ghaznavi2014b}). Therefore, simulations of this model suggest it would not yield the Hunt et al. \cite{Hunt2018} data shown in Figure \ref{Fig:HuntData-ALL}. A fundamental reason for this misalignment is that pore blocking increases transport limitations monotonically. During discharge, porosity only decreases and thus effective diffusion rates only increase. 

This fundamental modelling limitation can be observed throughout the modelling literature. A limited set of models have discharge simulations across multiple currents. Without a larger test set, as suggested in Cornish \& Marinescu \cite{cornish2022toward}, more models could not be assessed. Among the available information, several models did not produce any dip and recovery features \cite{Hofmann2014}, \cite{Ren2016}, \cite{Parke2020}, \cite{Parke2021}. From the models which did produce a dip and recovery feature, none produced a decrease in the feature with increasing current \cite{Zhang2015}, \cite{Thangavel2016}, \cite{Barai2016}, \cite{Zhang2016}, \cite{Andrei2018}; except for Danner \& Latz \cite{Danner2019}. Danner \& Latz utilise a sophisticated nucleation model which appears to adequately compete with the particle growth model and transport limitations to reduce the dip feature. However, further currents do not produce the large dip and recovery modelled in this work. 

Another secondary effect of precipitation dynamics is the covering and subsequent insulation of active surface area in the cathode. However, the change in relative active surface area from initial to final values requires orders of magnitude changes to produce a significant effect. From a modelling perspective, the active surface area parameter can be thought of as a multiplicative factor to the exchange current densities in the Butler-Volmer equation. Therefore, an investigation into the effect of alternative exchange current densities would provide an insight into the sensitivity of the voltage profile to active surface area effects. The work of Ghaznavi \& Chen \cite{Ghaznavi2014c} illustrates the point. In Figures 3 \& 4 of their work, the exchange current densities are altered by several orders of magnitude. To produce the strong changes in the discharge voltage profile observed for the 0.4C, 30$^{\circ}$C and 0.2C, 20$^{\circ}$C discharge profiles of Figure \ref{Fig:HuntData-ALL}, the active surface area would be required to go through changes of several orders of magnitude, which is atypical in the Kumaresan et al. \cite{Kumaresan2008} model and its subsequent modifications in the literature. A fortiori, the active surface area dynamic is plagued by the same fundamental issue as pore blocking: it is monotonic and therefore will not produce the temporary limitation required for the significant drop and recovery. 

\subsection{Transport Limitations}\label{sec:Transport Limitations}

The next most considered limitation of Lithium-Sulfur cells is the transport limitation. Indeed, a previous instantiation of the model presented in this work \cite{cornish2022toward} failed to produce the capacity-rate effect, as was easily observed through the suite of validation tests suggested in that work. A simple, pseudo-spatial model with constant diffusion can recover the capacity-rate effect (see Section \ref{sec:Supplementary Material}). Diffusion models that only consider the transport limitations associated with cathode materials can only provide a limited set of voltage data. In particular, a slow charge is required to produce the rate-capacity effect because cathode materials require time to transport outside of the cathode so that the subsequent discharge at a high current will not allow enough time for the species to transport back into the cathode and complete a full discharge. If the charge current is relatively high then most cathode material will remain in the cathode and so, although not all material will be utilised through the charging stages, the Coloumbic efficiency will be higher than experiments show for such operational conditions. This may be seen in, for example, the experiments of Mikhaylik \& Akridge \cite{Mikhaylik2004}. 

Even in the early work of Mikhaylik \& Akridge \cite{Mikhaylik2004}, the large dip and recovery can be observed with a significantly large salt concentration and a sufficiently high current (Figure 8, \cite{Mikhaylik2004}). If we assumed the transport limitations were solely due to cathodic materials, then Figure 10 of their work is difficult to explain. The figure shows the discharge voltage curves, all with the relatively high current of 350 mA. In cells of higher salt concentrations, this current was enough to produce the large voltage dip and recovery, but the cells in this particular figure had lower salt concentrations. The discharge curves were produced by previous charges of different currents. The slowest charge current, 20mA, induced shuttle and so the subsequent discharge high plateau capacity was very small but the low plateau capacity was approximately equal to the other discharges; the initial state of charge was lower but all materials had time to transport to and convert in the cathode. As the previous charge current is increased (curve 2 with 50mA  \& curve 3 with 200mA, Figure 10, \cite{Mikhaylik2004}) the subsequent discharge high plateau capacities increase. The low plateau capacities change by negligible amounts. Thus, the limiting features of these curves can be adequately explained by shuttle due to a low transport limitations during charge with the lower charging current. However, when the charge curve increases to 800mA (curve 5), the high plateau capacity is in between that of the 50mA and 200mA charge operations. This regression of the high plateau capacity cannot be explained by shuttle. Moreover, the discharge low plateau capacity is significantly lower than all other curves. Precipitation cannot explain this either, as the voltages are roughly equal to the other discharge curves, so nucleation rates should be the same (see Fan et al. \cite{fan2015mechanism}). This discharge rates are the same, so the total time for particle growth is comparable. This experimental condition was reproduced with the model presented in this work (section \ref{sec:Results & Discussion}).  

\subsection{Concentration Dependent Diffusion}

The missing component of the transport limitation which allows for the capture of the data in Figure \ref{Fig:HuntData-ALL} is a concentration dependent diffusion model. As more Lithium transports to the cathode, the total concentration of both the entire cell and, in particular, the cathode increases. This causes the effective diffusion of all species and, in particular, Lithium to decrease. This theory is supported by experimental evidence. As shown in Boenke et al.\cite{boenke2023role}, dependent upon the electrolyte the inclusion of polysulfide concentration can decrease the ionic conductivity by at least an order of magnitude. Moreover, the electrolyte viscosity can also increase by several orders of magnitude. It is reasonable to expect that the higher the concentration the more conductivity and viscosity are affected. Kirchoff et al. \cite{kirchhoff2023evaluation} showed that increasing the salt concentration in the solvent will yield the same large dip observed in the 0.4C, 30$^{\circ}$C and 0.2C, 20$^{\circ}$C discharge profiles of Figure \ref{Fig:HuntData-ALL}. This work will lend credence to the hypothesis that this increase in salt concentration produces a heightened sensitivity of the effective diffusion to the active species concentration. The dip and recovery becomes more pronounced at higher salt concentrations because the effective diffusion is lowered with the increase in concentration of any dissolved species. To drive the point home, the significant dip and recovery can also be observed when the E/S ratio is reduced significantly, as shown by Fan et al.  \cite{fan2017electrodeposition}. The work of Hunt et al. \cite{Hunt2018} will be used in conjunction with modelling to further these results by showing that this effect is significantly affected by temperature. It is known that viscosity increases with decreasing temperature \cite{gupta2020influence}. Therefore, altogether we may hypothesise that decreasing the temperature decreases the baseline diffusion coefficient, as a simple Arrhenius relation would suggest, and therefore makes the dynamic diffusion more sensitive to Lithium concentration in the same way that increasing the salt concentration of decreasing the E/S ratio would. 

The work of Danner \& Latz \cite{Danner2019} utilised a concentration based diffusion model. Their diffusion model (see Section \ref{sec:Supplementary Material}) utilised a viscosity model whereby increases in the total concentration of Sulfur species increased diffusion and subsequently lowered the effective diffusion rate. The model showed that, along side a sophisticated nucleation and particle growth model, higher discharge currents lead to a flattening and ultimate disappearance of the dip and recovery feature between the high and low voltage plateau (Figure 2a, \cite{Danner2019}). The reason why higher currents did not produce the re-emergence of a larger dip and recovery is due to the lack of Lithium in both the viscosity model and the electrochemistry. 

Many models utilise Lithium in the chemical reactions \cite{Kumaresan2008,Neidhardt2012, Fronczek2013, Hofmann2014, Danner2015, Zhang2015, Zhang2016, Thangavel2016, Danner2019}. However, as previously stated the precipitation mechanisms cannot explain the Hunt et al. \cite{Hunt2018} data. Therefore, transport limitations of Lithium affects precipitation dynamics but will not cause the large dip and recovery effect. Ren et al. \cite{Ren2016} utilise Lithium as part of the cathode electrochemical pathway, however it is only part of the final reaction, coupling with the nucleation model. Dynamic diffusion based on Lithium concentration is included, but the electrochemical pathways does not utilise Lithium until the final reaction. Inclusion of Lithium in the Butler-Volmer equations, as done in the model presented in this work, allows the Lithium transport limitations to have the desired effect. Therefore, the inclusion of Lithium in the modelled electrochemistry and transport limitations leads to more valid results.

In this work, a simple model for the dynamic diffusion mechanism is presented to show that the desired affects are qualitatively achievable. 

\section{Model Development}\label{sec:Model Development}
A zero dimensional model is insufficient to capture the capacity-rate effect. This was exhibited through the validation test set in the model of Cornish \& Marinescu \cite{cornish2022toward}, and is expected for all similar models. However, a 1D model is not necessary. To obtain the capacity-rate effect, it is sufficient to allow transport between two regions: the cathode region where electrochemical reactions take place and the separator region where no electrochemical reactions occur. This is demonstrated in section \ref{sec:Supplementary Material}, where a simple pseudo-spatial model is constructed by coupling two 0D models for the separator and cathode regions.

However, as discussed in section \ref{sec:Dynamic Limitations}, the correct behaviour across different relative charge and discharge currents requires that the electrochemistry is dependent upon not only Sulfur species, as every model considers, but also on Lithium. Therefore, the electrochemical pathways used in this work is the 3-stage reaction,

\begin{eqnarray*}
	S_8^0 + 4Li^+ + 4e^- \to & 2Li_2S_4 \\
	Li_2S_4 + 2Li^+ + 2e^- \to & 2Li_2S_2 \\
	Li_2S_2 + 2Li^+ + 2e^- \to & 2Li_2S \downarrow \\  
\end{eqnarray*}

It is exceedingly important to ensure free Lithium cations are required as reactants but not products. Otherwise, the concentration of Lithium cations acts as a multiplicative factor that modulates the entire Butler-Volmer equation in such a way that higher concentrations decrease over-potential, which is exactly opposite of our hypothesis. 

Given this electrochemistry, the cathode region dynamics are modelled as a zero-dimensional model with an exchange term for species to transport to the separator region. The cathode region model also includes a slow-nucleation particle growth model for the precipitated species. 

\begin{eqnarray*}
	\frac{dC_{S_8}^{cat}}{dt}  & = & -\frac{i_H}{n_H\nu F} - D_8(C_{S_8}^{cat} - C_{S_8}^{sep}) \\
	\frac{dC_{Li_2S_4}^{cat}}{dt} &= & \frac{2i_H}{n_H \nu F} -\frac{i_M}{n_M\nu F} - D_4(C_{Li_2S_4}^{cat} - C_{Li_2S_4}^{sep}) \\
	\frac{dC_{Li_2S_2}^{cat}}{dt} &= & \frac{2i_M}{n_M \nu F} -\frac{i_L}{n_L\nu F} - D_2(C_{Li_2S_2}^{cat} - C_{Li_2S_2}^{sep}) \\
	\frac{dC_{Li_2S}^{cat}}{dt} &= & \frac{2i_L}{n_L \nu F} -\frac{M_s k_p }{\rho_s}C_{S_p}^{cat}(C_{Li_2S}^{cat}- K_{sp}) \\
	& & - D_1(C_{Li_2S}^{cat} - C_{Li_2S}^{sep}) \\
	\frac{dC_{Sp}^{cat}}{dt} &= & \frac{M_s k_p }{\rho_s}C_{S_p}^{cat}(C_{Li_2S}^{cat}- K_{sp}) \\
	\frac{dC_{Li^+}^{cat}}{dt} & = & - D_{Li^+}(C_{Li^+}^{cat} - C_{Li^+}^{sep}) - \frac{I}{\nu F}
\end{eqnarray*}

which includes the current condition, 

\begin{equation}
	I = i_H + i_M + i_L
\end{equation}

for
\begin{eqnarray*}
	i_H & = & a j_H^0\bigg( (C_{Li^+}^{cat})^4C_{S8}^{cat}e^{-\frac{n_HF}{2RT}(V-E_H^0)} \\
	& & - (C_{Li_2S_4}^{cat})^2e^{\frac{n_HF}{2RT}(V-E_H^0)} \bigg) \\
	i_M & = & a j_M^0\bigg( (C_{Li^+}^{cat})^2C_{Li_2S_4}^{cat}e^{-\frac{n_MF}{2RT}(V-E_M^0)} \\
	& & - (C_{Li_2S_2}^{cat})^2e^{\frac{n_MF}{2RT}(V-E_M^0)} \bigg) \\
	i_L & = & a j_L^0\bigg( (C_{Li^+}^{cat})^2C_{Li_2S_2}^{cat}e^{-\frac{n_LF}{2RT}(V-E_L^0)} \\
	& & - (C_{Li_2S}^{cat})^2e^{\frac{n_LF}{2RT}(V-E_L^0)} \bigg) 
\end{eqnarray*}

The separator region is modelled as a zero-dimensional model with an exchange term for species to transport to the cathode. Also included is a simple conversion model between $S_8$ and $Li_2S_4$ to allow self-discharge and, along with the transport, the shuttle phenomena. The dynamic equations are,  

\begin{eqnarray*}
	\frac{dC_{S_8}^{sep}}{dt}  & = & -k_sC_{S_8}^{sep} + D_8(C_{S_8}^{cat} - C_{S_8}^{sep}) \\
	\frac{dC_{Li_2S_4}^{sep}}{dt} &= & 2k_sC_{S_8}^{sep} + D_4(C_{Li_2S_4}^{cat} - C_{Li_2S_4}^{sep}) \\
	\frac{dC_{Li_2S_2}^{sep}}{dt} &= & D_2(C_{Li_2S_2}^{cat} - C_{Li_2S_2}^{sep}) \\
	\frac{dC_{Li_2S}^{sep}}{dt} &= & D_1(C_{Li_2S}^{cat} - C_{Li_2S}^{sep}) \\
	\frac{dC_{Li^+}^{cat}}{dt} & = & D_{Li^+}(C_{Li^+}^{cat} - C_{Li^+}^{sep}) + \frac{I}{\nu F}
\end{eqnarray*}

We should note that the shuttle model assumes that $S_8$ transforms on the anode surface and therefore the subsequent creation of $Li_2S_4$ does not require free ions, rather the Lithium is supplied directly from the anode material.

Finally, in order to capture the dynamic concentration based resistance, the diffusion coefficients are all modulated by the total concentration of dissolved species. The simple expression given below is a first order approximation of the concentration based diffusion model given in Danner \& Latz \cite{Danner2019} (see section \ref{sec:Supplementary Material} for derivation).

\begin{eqnarray*}
	D_8 =  \frac{D_8^0}{1 + \gamma C_T}, & D_4 =  \frac{D_4^0}{1 + \gamma C_T}, & D_2 =  \frac{D_2^0}{1 + \gamma C_T}, \\
	D_1 =  \frac{D_1^0}{1 + \gamma C_T}, & D_{Li^+} =  \frac{D_{Li^+}^0}{1 + \gamma C_T}, & 
\end{eqnarray*}

where $\gamma$ is a fitted constant and, 

\begin{equation}
	C_T = \sum_{i \in \{sep, cat\}}C_{S_8}^{i} + C_{Li_2S_4}^{i}+ C_{Li_2S_2}^{i}+ C_{Li_2S}^{i} + C_{Li^+}^{i}
\end{equation}

\section{Model Validation}\label{sec:Model Validation}
The parameterisation of the model began with fitting the initial conditions and all parameters to the 40$^\circ$C, 0.2C discharge voltage curve shown in Figure \ref{Fig:HuntData-ALL}. The value of $\gamma$, the sensitivity of the diffusion to total dissolved species concentration, was fit such that the 40$^\circ$C, 0.8C discharge voltage matched the 20$^\circ$C, 0.2C discharge voltage curve. This was to ensure the 40$^\circ$C dynamics could produce the behaviour observed in the 30$^\circ$C, 0.4C and 20$^\circ$C, 0.2C voltage curves. Therefore, the 40$^\circ$C, 0.4C and 40$^\circ$C, 1.6C discharge voltage curves are technically out-of-sample. The entire parameterisation procedure above was performed for the 5th discharge after a 0.2C-0.1C discharge-charge cycle, reflecting the experimental operation. The shuttle constant was the only value fit to the 40$^\circ$C, 0.1C charge. 

The parameterisation of the temperature dependent parameters (see Section \ref{sec:Supplementary Material}) was performed similarly. In particular, the parameters at 30$^\circ$C were found by fitting the 0.2C and 0.4C discharge voltage curves. The parameters at 30$^\circ$C were found by fitting the 0.2C and 0.4C discharge voltage curves. Shuttle was also fit for each temperature's charge curve. All other temperature-current operational conditions constitute out-of-sample predictions. 

As discussed in Cornish \& Marinescu \cite{cornish2022toward}, many other operational conditions are available for testing the model to compare the model predictions to those of other models and also to common experimental results. These predictions, along with the full spectrum of discharge voltage curves associated with the Hunt et al. \cite{Hunt2018} data, are given in Section \ref{sec:Results & Discussion} with a discussion of the meaning of these plots. 

\section{Model Results \& Discussion}\label{sec:Results & Discussion}

\begin{figure}[h]
\includegraphics[width=\columnwidth]{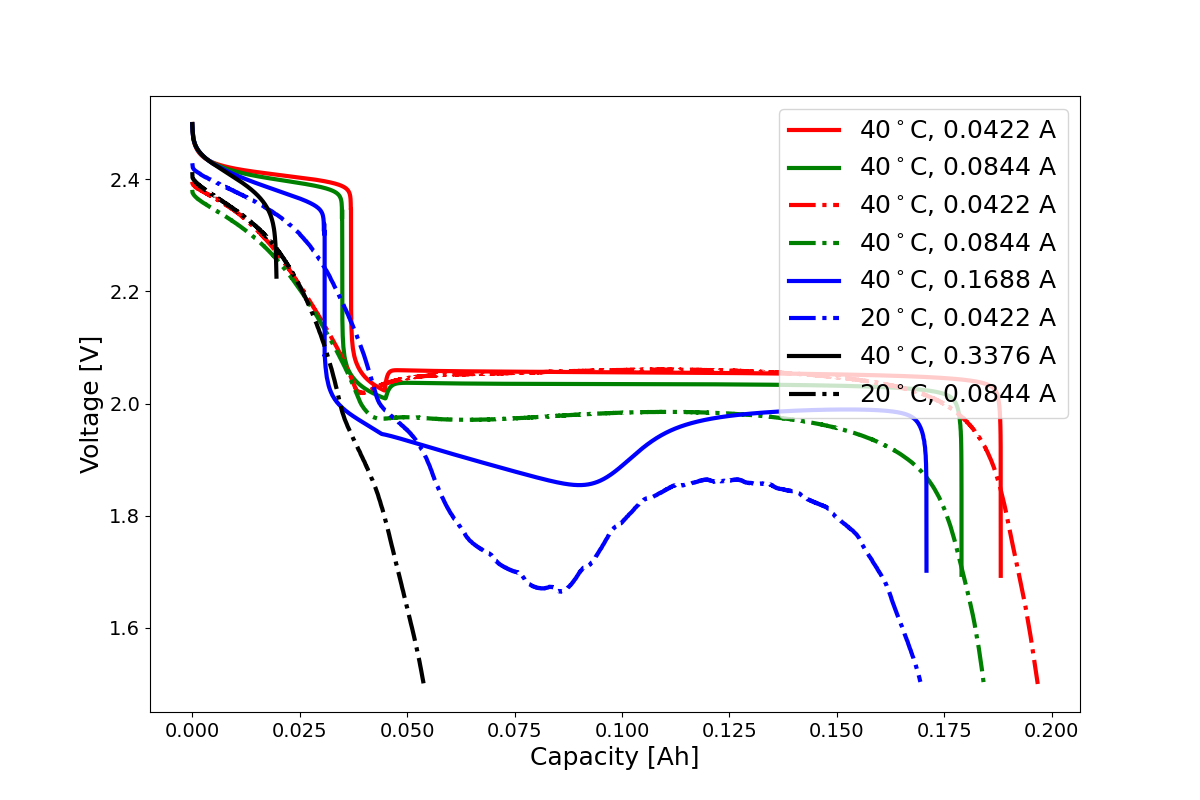}
\caption{Simulated discharge voltage curves at 40$^\circ$C. Range of currents exceed that of the experiment of Hunt et al. \cite{Hunt2018}, so the comparison is made to the relevant data through use of the Current-Temperature Isometry. Solid lines represent the simulations while dashed lines represent the experimental data.}
\label{40_discharge_all}
\end{figure}

Figure \ref{40_discharge_all} displays the 40$^\circ$C temperature discharge curves compared to the Hunt et al. \cite{Hunt2018} data. The lower discharge current matches well, given the simplicity of the model. The higher discharge current yields a lower low plateau voltage as well as the expected capacity rate effect. The lower sensitivity of the lower plateau voltage to increases in current may be due to a relatively small transport limitation effect. Such an effect is far more pronounced at higher currents (see below). Nonetheless, a qualitative match to the basic features of current-dependence is observed against the experimental data. Moreover, given the currents utilised, a decrease in the voltage dip can be seen to emerge. 

\begin{figure}[h]
\includegraphics[width=\columnwidth]{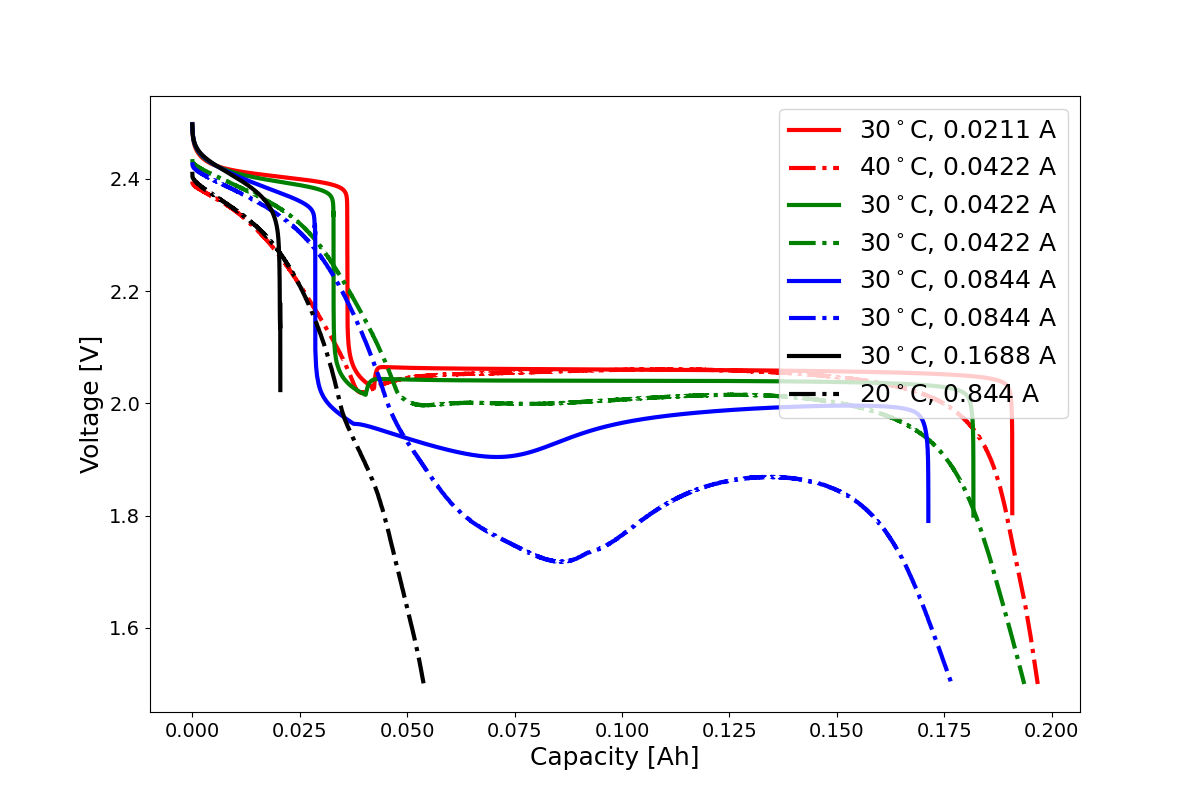}
\caption{Simulated discharge voltage curves at 30$^\circ$C. Range of currents exceed that of the experiment of Hunt et al. \cite{Hunt2018}, so the comparison is made to the relevant data through use of the Current-Temperature Isometry. Solid lines represent the simulations while dashed lines represent the experimental data.}
\label{30_discharge_all}
\end{figure}

\begin{figure}[h]
\includegraphics[width=\columnwidth]{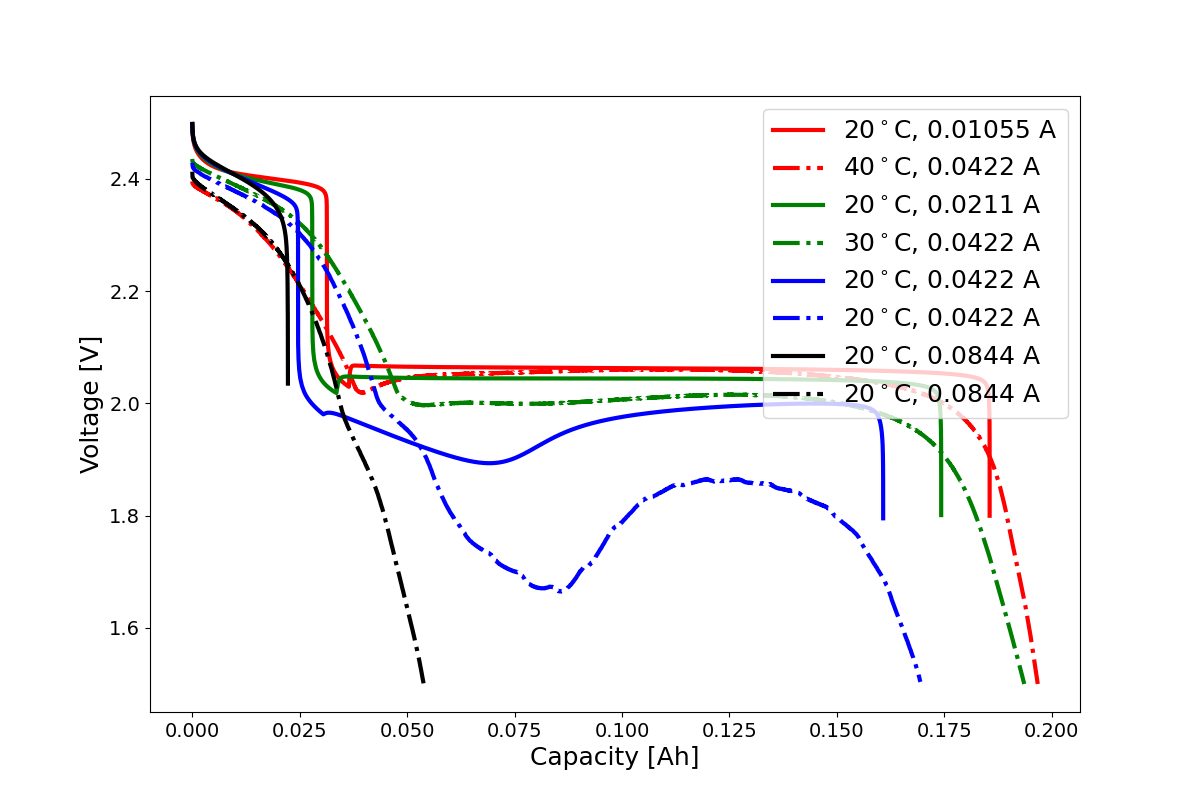}
\caption{Simulated discharge voltage curves at 20$^\circ$C. Range of currents exceed that of the experiment of Hunt et al. \cite{Hunt2018}, so the comparison is made to the relevant data through use of the Current-Temperature Isometry. Solid lines represent the simulations while dashed lines represent the experimental data.}
\label{20_discharge_all}
\end{figure}

Figure \ref{30_discharge_all} displays the 30$^\circ$C temperature discharge curves and Figure \ref{20_discharge_all} displays the 20$^\circ$C temperature discharge curves. Both are directly compared to the Hunt et al. \cite{Hunt2018} data. The pronounced voltage dip and recovery can now be easily seen in the experimental data and the simulation. The lower currents obtain the more regular LiS voltage profile. At moderate and high currents the main feature of this work is exhibited. The highest current would have a voltage that drops to the lower voltage limit, except numerical error emerges. The concentration of Lithium in the cathode becomes too close to zero to allow further resolution in time of the system. 

\begin{figure}[h]
\includegraphics[width=\columnwidth]{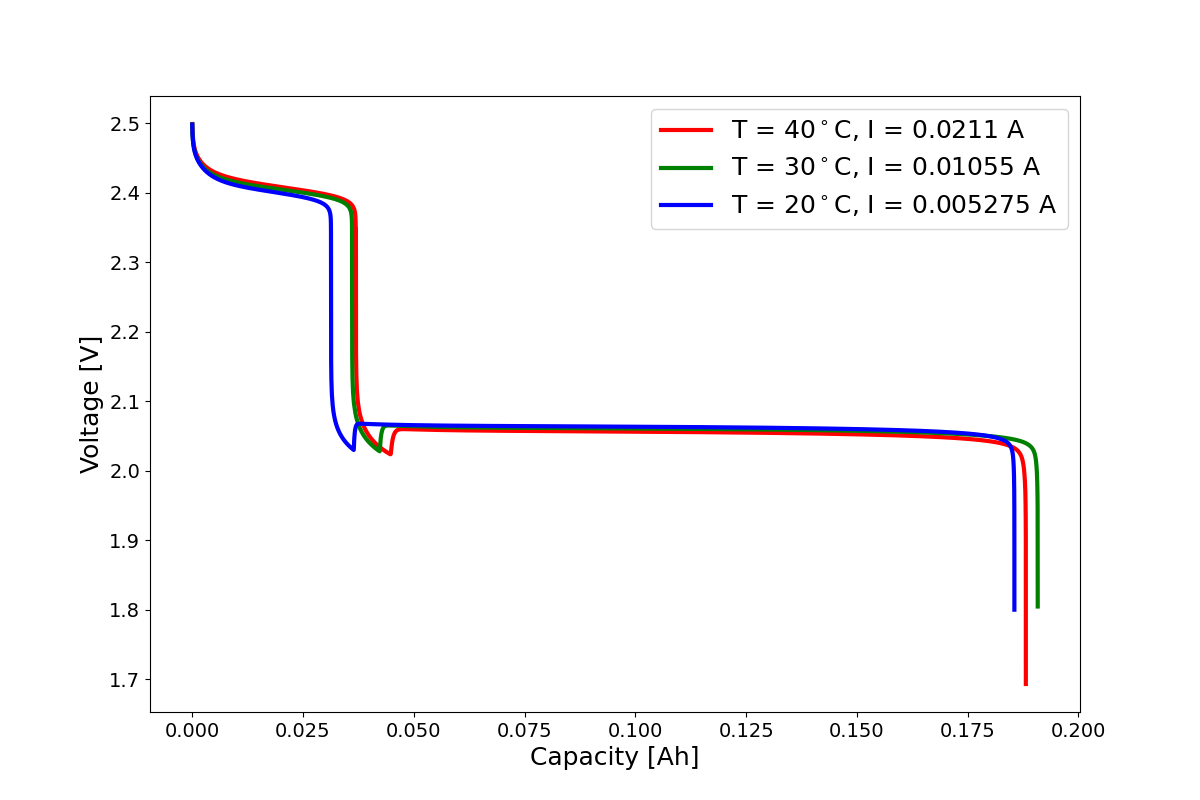}
\caption{Simulated discharge voltage Current-Temperature Isometry. The qualitative shape of the curves remains the same and most of the quantitative features all approximately equal.}
\label{Isometry_1}
\end{figure}

\begin{figure}[h]
\includegraphics[width=\columnwidth]{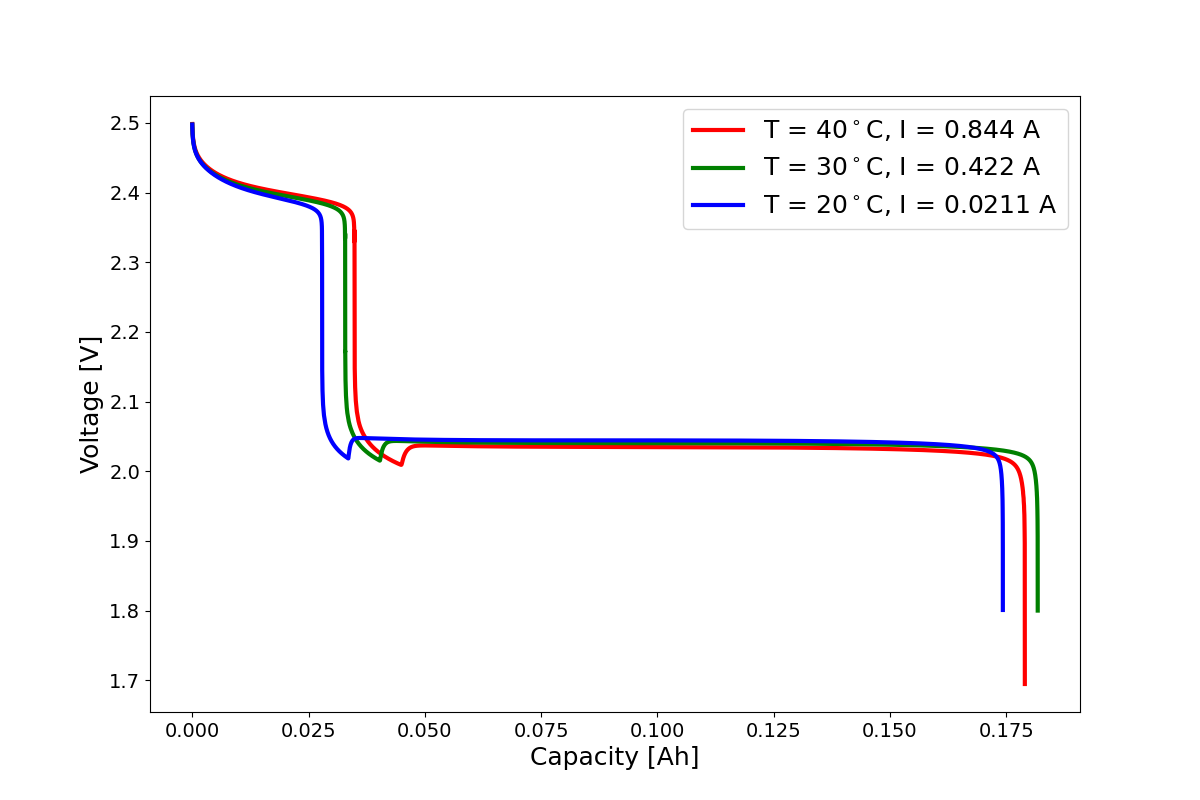}
\caption{Simulated discharge voltage Current-Temperature Isometry. The qualitative shape of the curves remains the same and most of the quantitative features all approximately equal.}
\label{Isometry_2}[h]
\end{figure}

\begin{figure}[h]
\includegraphics[width=\columnwidth]{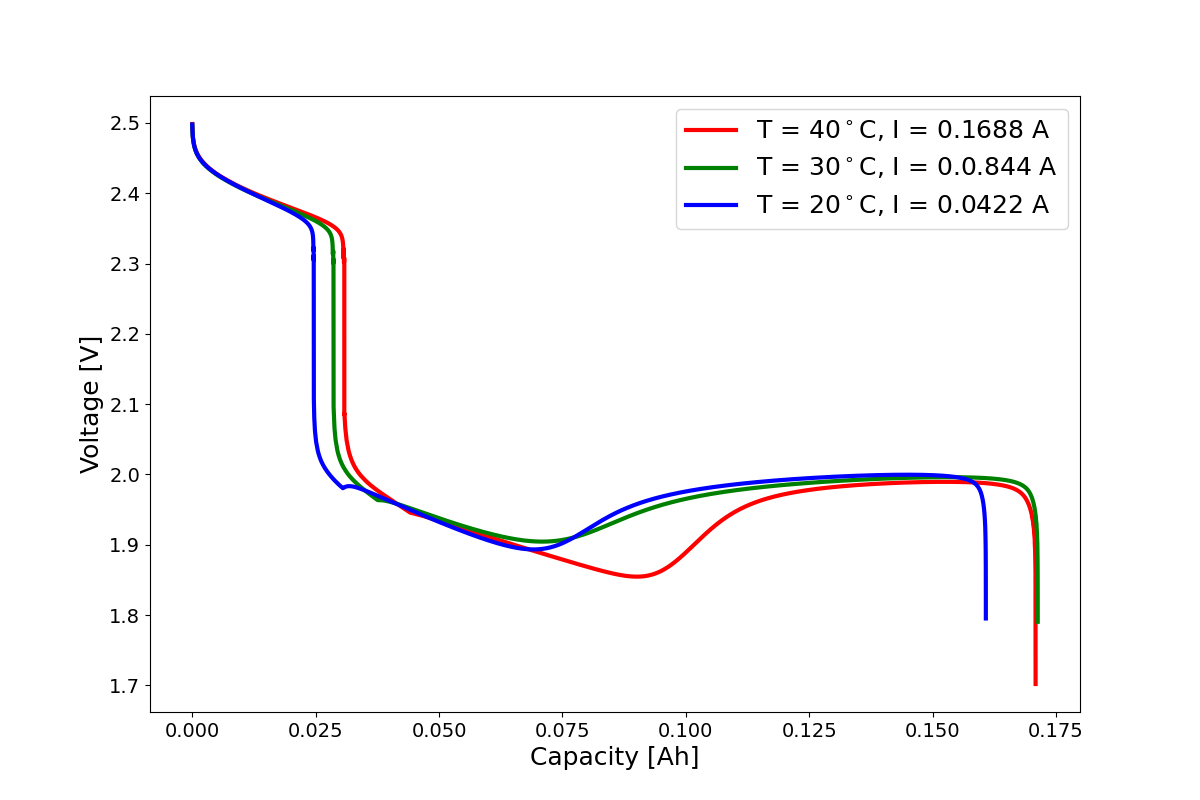}
\caption{Simulated discharge voltage Current-Temperature Isometry. The qualitative shape of the curves remains the same and most of the quantitative features all approximately equal.}
\label{Isometry_3}
\end{figure}

\begin{figure}[h]
\includegraphics[width=\columnwidth]{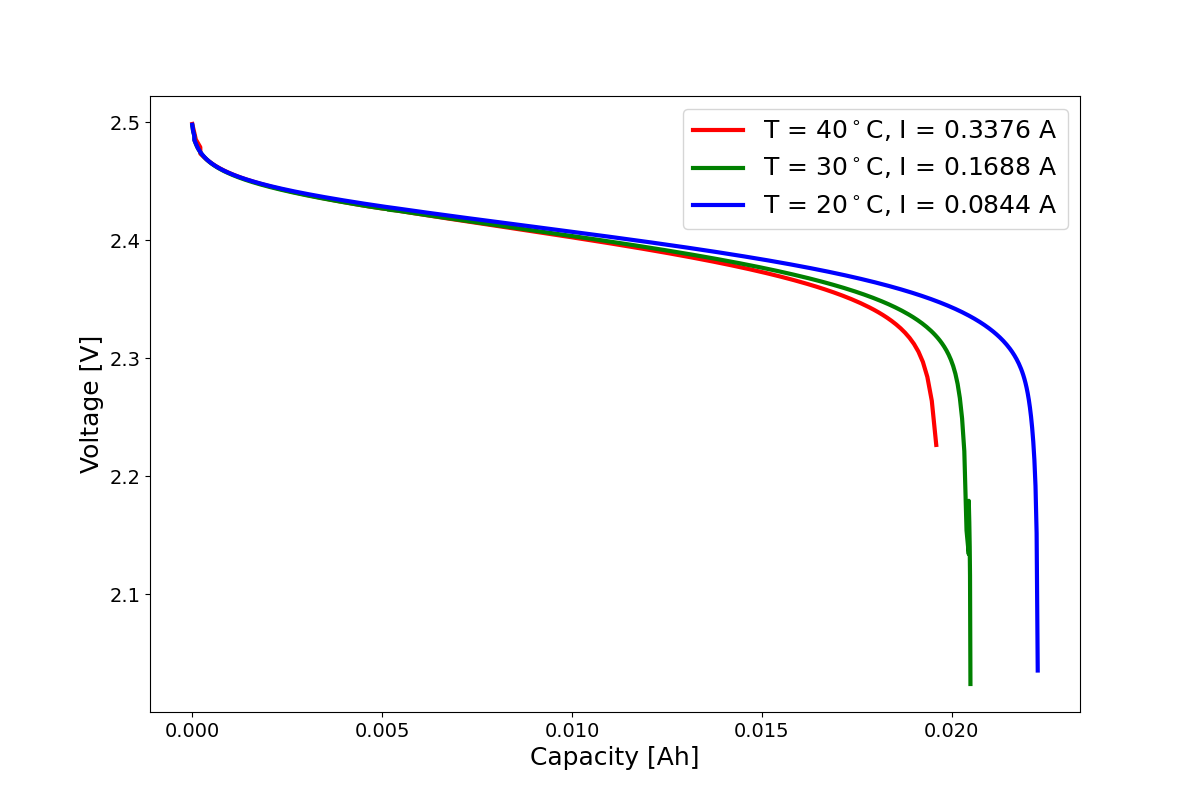}
\caption{Simulated discharge voltage Current-Temperature Isometry. The qualitative shape of the curves remains the same and most of the quantitative features all approximately equal.}
\label{Isometry_4}
\end{figure}

Figures \ref{Isometry_1}, \ref{Isometry_2}, \ref{Isometry_3}, \& \ref{Isometry_4} displays the Current-Temperature Isometry via the discharge voltage curves from all three temperature conditions. It is clear that, excusing state-of-charge issues due to shuttle, the Current-Temperature Isometry is exhibited by the simulation data. Unfortunately, experimental data is limited. Added to each plot is the available data for the Isometry in question. 

\begin{figure}[h]
\includegraphics[width=\columnwidth]{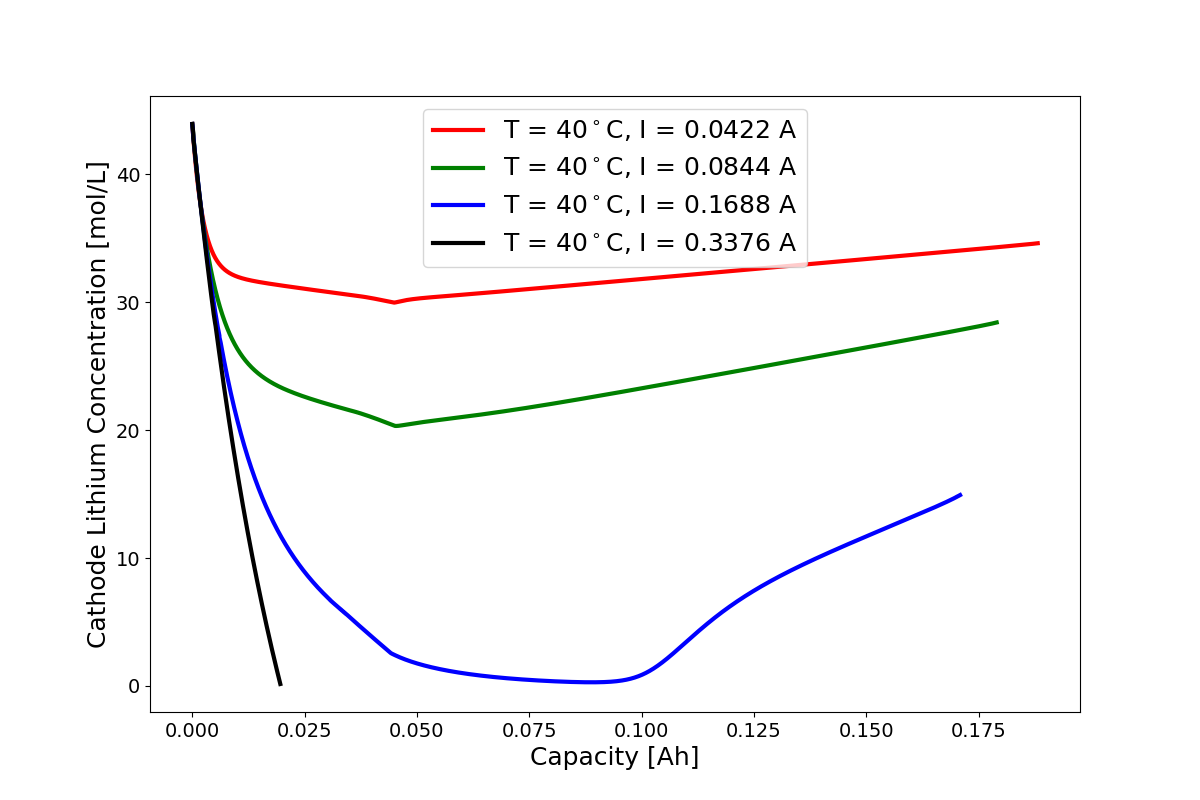}
\caption{Simulated discharge Lithium cation concentrations in the cathode. The values close to zero cause the voltage to drop significantly. The 0.3376 A discharge causes the concentration to drop to zero, setting the cell voltage below operating limits. The 0.1688 A discharge allows the concentrations to remain close to zero, causing the severe but temporary voltage drop. }
\label{40_Lithium_concentrations}
\end{figure}

Figure \ref{40_Lithium_concentrations} displays the cause of the Current-Temperature Isometry via the concentration of Lithium cations during discharge. The plot shows the 40$^\circ$C simulation data. The other temperatures have much the same plots. It is clear that the severe dip in voltage is due to the low supply of free Lithium in the cathode. As Lithium cations are collected in the $Li_2S_n$ species and are not sufficiently resupplied in the cathode from the anode, the Butler-Volmer equation becomes unbalanced in the discharge direction. The over-potential therefore increases greatly.

\begin{figure}[h]
\includegraphics[width=\columnwidth]{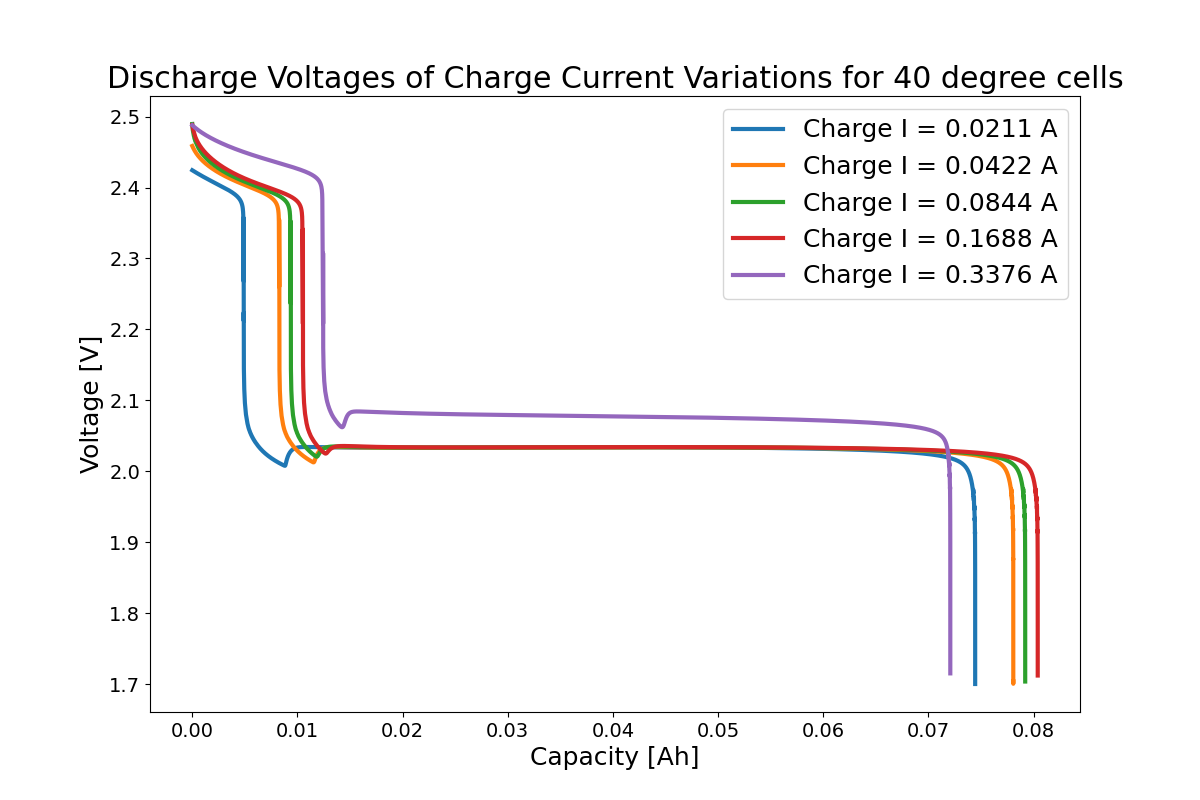}
\caption{Simulated effect of transport limitations due to previous charge current. Low charge currents induce shuttle, which decreases both the overall state-of-charge as well as the high plateau capacity. Larger currents reduce the shuttle effect. The highest current induces the Lithium cation transport limitation.}
\label{charge_current_variations}
\end{figure}

Figure 10 in the work of Mikhaylik \& Akridge \cite{Mikhaylik2004} displays experimental results where a relatively high discharge current was utilised subsequent to various charge currents. The interpretation of such results in section \ref{sec:Transport Limitations} above was that lower discharge currents led to more shuttle and hence lower overall discharge capacity. As the charge current increases, the shuttle effect decreases and  resultant discharge has a higher capacity. This effect remains until another transport limitation occurs: Lithium transport. The highest current led to a decrease in high plateau capacity as well as low plateau capacity. This cannot be explained by cathodic material transport limitations. It may be possible that the highest charge current left much precipitated material in the cathode and so the subsequent discharge was limited by this precipitation history effect. However, Figure 7 from Mikhaylik \& Akridge \cite{Mikhaylik2004} shows that transport limitations will lead to similar high plateaus capacities but smaller low plateau capacities (100 \& 350 mA curves, specifically). Therefore, it is likely the result of charging at a relatively high current is that Lithium cations do not have sufficient time to transport across the cell, leading to capacity loss. The out-of-sample predictions of the model presented here are displayed in Figure \ref{charge_current_variations}. Increasing the charge current leads first to higher discharge capacities due to shuttle but then leads to a transport limitation that decreases the discharge capacity. This is precisely the behaviour observed in the experiment.

The specific effects of Lithium cation transportation during charging are not the focus of this work. However, for completeness the charge voltage curves, where only the shuttle coefficient was fit to the data, are given in section \ref{sec:charge_ohmic}. This includes the effects of depth-of-discharge on the subsequent charge voltage kink effect. Moreover, the Ohmic resistance curves, where only two parameters are fit (see Cornish \& Marinescu \cite{cornish2022toward}), are also provided in section \ref{sec:charge_ohmic}.

\section{Conclusion}\label{sec:Conclusion}
The model presented in this work captures all features considered in the model validation standard suite of tests provided by Cornish \& Marinescu \cite{cornish2022toward}. The psuedo-spatial feature, along with Lithium included in the electrochemical pathways, was enough to provide transport effects that yields the completion of the feature suite. 

Beyond the basic validation tests, the model also captured a prominent feature in LiS batteries that no other model appears to have produced or possibly could produce. The pronounced dip and recovery at high currents provides evidence for transport limitations due to Lithium that strongly affects cell performance.

The generalisation of the experimental data suggests a new relationship for LiS batteries: the Current-Temperature Isometry. There appears to be a direct relationship between temperature and current: increasing temperature linearly and current exponentially will maintain a similar performance to lower temperature and current cells, as observed from discharge profile. It is not clear at this point how this Isometry may affect degradation mechanisms and therefore long term behaviour. However, for single cycles it appears this feature is controlled through Li cation transport limitations. 

A corollary to the Li cation transport limitation is the change in high and low discharge capacities after relatively high charge rates. This phenomena was shown early in the modelling literature through the modelling and experimental work of Mikhaylik \& Akridge \cite{Mikhaylik2004}. This phenomena can now be explained by the Li cation transport limitation. 

As a final point, we hope the modelling community can observe through this work and Cornish \& Marinescu \cite{cornish2022toward} the value of the validation test set. The model presented above was born out of a detailed investigation into the causes for the flaws of the model from Cornish \& Marinescu \cite{cornish2022toward}. That investigation could not have occurred without the model validation suite. 

\section{Supplementary Material}\label{sec:Supplementary Material}

\subsection{Dynamic Diffusion Model Derivation}
The concentration based diffusion model from Danner \& Latz \cite{Danner2019} is, 

\begin{equation}
	D_i^{\textit{eff}} = D_i^0 \varepsilon_{\textit{elyte}}^{\beta}\frac{\eta_{\textit{elyte}}^0}{\eta_{\textit{elyte}}}
\end{equation}

where $D_i^0$ is a constant, $\varepsilon_{\textit{elyte}}$ is the porosity of the pore network, $\beta$ a constant, $\eta_{\textit{elyte}}^0$ a constant representing the electrolyte viscosity at the initial salt concentration, and $\eta_{\textit{elyte}}$ is defined as, 

\begin{equation}
	\eta_{\textit{elyte}} = \alpha e^{\gamma c_S}
\end{equation}
 
 for parameters $\alpha$, $\gamma >0$ set to the experimental work of Fan et al. \cite{Fan2016}, and where $c_S$ is the total Sulfur concentration. To derive the concentration based model defined in section \ref{sec:Model Development}, we first assume porosity is constant and absorb its constant into the diffusion constant. Then, we utilise total dissolved active species, $C_T$ instead of only the Sulfur species. Next, we expand the electrolyte function as the linear portion of a Taylors Series, 
 
 \begin{eqnarray}
 	\eta_{\textit{elyte}}(C_T) & \approx &  \eta_{\textit{elyte}}(0) + C_T\gamma \eta_{\textit{elyte}}(0) \\
 	& = & \eta_{\textit{elyte}}^0 + C_T\gamma\eta_{\textit{elyte}}^0 
 \end{eqnarray}
 
Therefore,
 
 \begin{eqnarray}
 	\frac{\eta_{\textit{elyte}}^0}{\eta_{\textit{elyte}}} & \approx & \frac{\eta_{\textit{elyte}}^0}{\eta_{\textit{elyte}}^0 + C_T\gamma\eta_{\textit{elyte}}^0 } \\
 	& = & \frac{1}{1 + C_T\gamma}
 \end{eqnarray}
 
 \subsection{Constant Diffusion Effects}
 Below we show the effects of the discharge voltage dip-and-recovery given a constant diffusion coefficient. By setting $\gamma = 0$, the effective diffusion coefficient becomes constant. Therefore, there is no transport resistance that rises to a peak during the plateau transition, and so the experimentally observed effects are not able to be captured. However, the rate-capacity effect can be captured under specified charging regimes. 
 
\begin{figure}[h]
\includegraphics[width=\columnwidth]{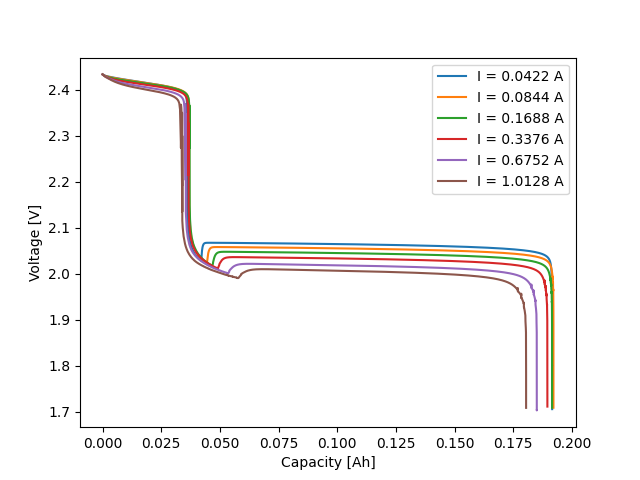}
\caption{Simulations with a constant diffusion coefficient.}
\label{static_diffusion}
\end{figure}

 Given an initial condition found through a charge-discharge cycle repeated four times under the experimental operational condition, the change in capacity is displayed in Figure \ref{static_diffusion}. Additionally, larger discharge currents were performed to ensure the capacity-rate effect would be more pronounced under the same parameterisation as the main model of this paper. As shown, the voltage dip cannot be enlarged after first decreasing. The rate-capacity effect can be recovered, as is expect from transport models. 
 
 \subsection{Lithium \& Electrochemical Reactions}
 We claim that it is a requirement that lithium is not only included in the cathodic electrochemical reactions, but also that it is to be included in the forward reaction term (from the perspective of discharge).

\begin{figure}[h]
\includegraphics[width=\columnwidth]{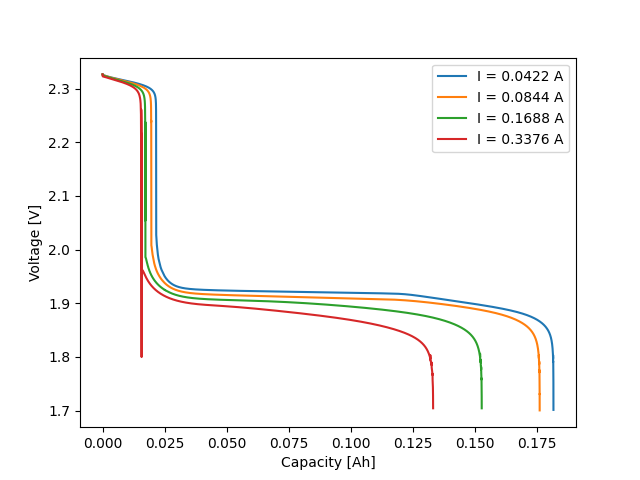}
\caption{Simulations with electrochemistry that does not include the Lithium cations.}
\label{Electrochemistry_1}
\end{figure} 

 If there is no Lithium in the electrochemical model, then the dynamic diffusion model presented in this work does not yield the desired results. Figure \ref{Electrochemistry_1} shows just this case. The only modifications to the model are the removal of Lithium from the Butler-Volmer equations to reflect the following electrochemistry,
 
 \begin{eqnarray*}
	S_8^0 + 4e^- \to & 2S_4^{2-} \\
	S_4^{2-} + 2e^- \to & 2S_2^{2-} \\
	S_2^{2-} + 2e^- \to & 2S^{2-} \downarrow \\  
\end{eqnarray*}

This yields the Butler-Volmer equations, 

\begin{eqnarray*}
	i_H & = & a j_H^0\bigg( C_{S8}^{cat}e^{-\frac{n_HF}{2RT}(V-E_H^0)} \\
	& & - (C_{S_4}^{cat})^2e^{\frac{n_HF}{2RT}(V-E_H^0)} \bigg) \\
	i_M & = & a j_M^0\bigg(C_{S_4}^{cat}e^{-\frac{n_MF}{2RT}(V-E_M^0)} \\ 
	& & - (C_{S_2}^{cat})^2e^{\frac{n_MF}{2RT}(V-E_M^0)} \bigg) \\
	i_L & = & a j_L^0\bigg( C_{S_2}^{cat}e^{-\frac{n_LF}{2RT}(V-E_L^0)} \\
	& & - (C_{S}^{cat})^2e^{\frac{n_LF}{2RT}(V-E_L^0)} \bigg) 
\end{eqnarray*}

The Lithium may exit the electrolyte through the chemical reaction with $S^{2-}$ as precipitated $Li_2S$. Therefore, the other necessary modification is for the precipitation model. The only dynamic equations which require change are 
 
 \begin{eqnarray*}
	\frac{dC_{S}^{cat}}{dt} &= & \frac{2i_L}{n_L \nu F} -\frac{M_s k_p }{\rho_s}C_{S_p}^{cat}((C_{Li^+}^{cat})^2C_{S}^{cat}- K_{sp}) \\
	& &  - D_1(C_{S}^{cat} - C_{S}^{sep}) \\
	\frac{dC_{Sp}^{cat}}{dt} &= & \frac{M_s k_p }{\rho_s}C_{S_p}^{cat}((C_{Li^+}^{cat})^2C_{Li_2S}^{cat}- K_{sp}) \\
	\frac{dC_{Li^+}^{cat}}{dt} & = & - D_{Li^+}(C_{Li^+}^{cat} - C_{Li^+}^{sep}) \\ 
	& & - 2\frac{M_s k_p }{\rho_s}C_{S_p}^{cat}((C_{Li^+}^{cat})^2C_{Li_2S}^{cat}- K_{sp})
	\end{eqnarray*}
	
The precipitation rate was increased by an order of magnitude to account for the precipitation model reformulation.

An alternative model could express the requirement of Lithium for the electrochemical reactions, but that the Lithium is only a catalyst. Consequently, the electrochemical equations can be expressed as,
 
\begin{eqnarray*}
	S_8^0 + 4e^- + 4Li^+ \to & 2S_4^{2-} + 4Li^+ \\
	S_4^{2-} + 2e^- + 2Li^+\to & 2S_2^{2-}+ 2Li^+ \\
	S_2^{2-} + 2e^- + 2Li^+ \to & 2S^{2-}+ 2Li^+ \downarrow \\  
\end{eqnarray*}
 
 That is to say that the Butler-Volmer equation is 
 
 \begin{eqnarray*}
	i_H & = & a j_H^0(C_{Li^+}^{cat})^4\bigg( C_{S8}^{cat}e^{-\frac{n_HF}{2RT}(V-E_H^0)} \\
	& & - (C_{S_4}^{cat})^2e^{\frac{n_HF}{2RT}(V-E_H^0)} \bigg) \\
	i_M & = & a j_M^0(C_{Li^+}^{cat})^2\bigg( C_{S_4}^{cat}e^{-\frac{n_MF}{2RT}(V-E_M^0)} \\
	& & - (C_{S_2}^{cat})^2e^{\frac{n_MF}{2RT}(V-E_M^0)} \bigg) \\
	i_L & = & a j_L^0(C_{Li^+}^{cat})^2\bigg( C_{S_2}^{cat}e^{-\frac{n_LF}{2RT}(V-E_L^0)} \\
	& & - (C_{S}^{cat})^2e^{\frac{n_LF}{2RT}(V-E_L^0)} \bigg) 
\end{eqnarray*}

\begin{figure}[h]
\includegraphics[width=\columnwidth]{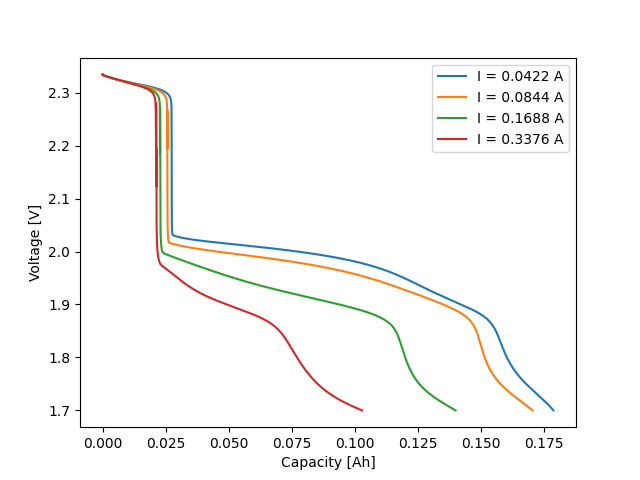}
\caption{Simulations with electrochemistry that utilises the Lithium cations as non-transforming species required for the reaction.}
\label{Electrochemistry_2}
\end{figure}

The results of this model are given in Figures \ref{Electrochemistry_2}. It is clear that not only do the discharge plots not appear similar to the standard discharge voltage curves due to the extra transition and plateau regions, but also that these simulations lack the severe voltage drop and recovery concerning this work. 

\subsection{Charge and Ohmic Resistance Simulations}\label{sec:charge_ohmic}
 
\begin{figure}[h]
\includegraphics[width=\columnwidth]{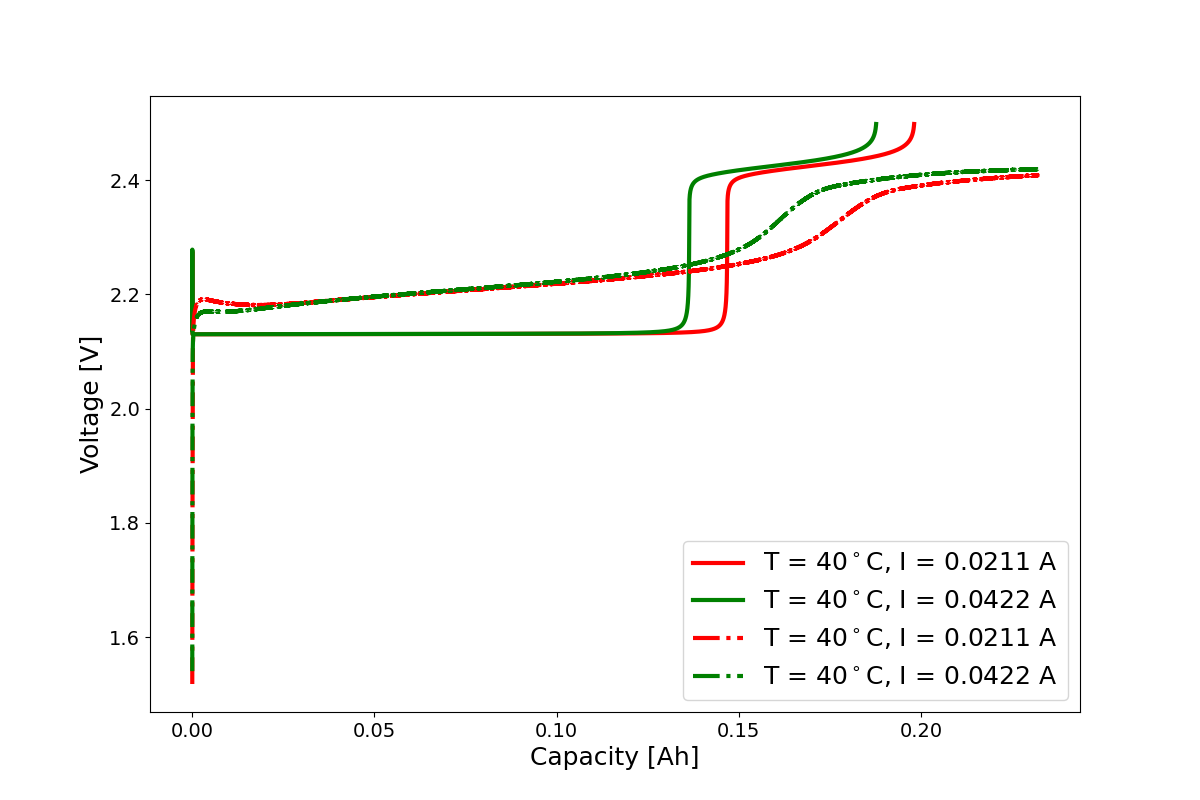}
\caption{Simulated effect charge for 40$^\circ$C cell. Experimental data is represented by dashed lines while solid lines represent the simulated data.}
\label{40_charge}
\end{figure} 

\begin{figure}[h]
\includegraphics[width=\columnwidth]{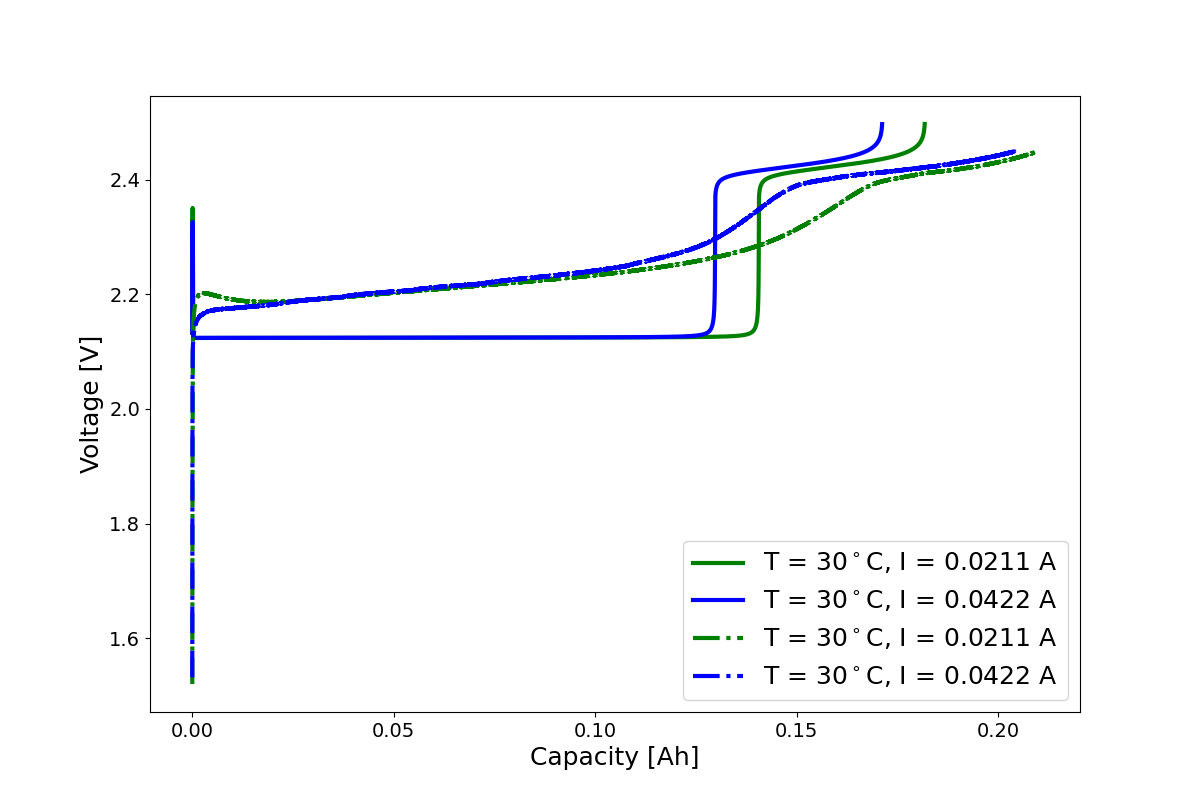}
\caption{Simulated effect charge for 30$^\circ$C cell. Experimental data is represented by dashed lines while solid lines represent the simulated data.}
\label{30_charge}
\end{figure} 

\begin{figure}[h]
\includegraphics[width=\columnwidth]{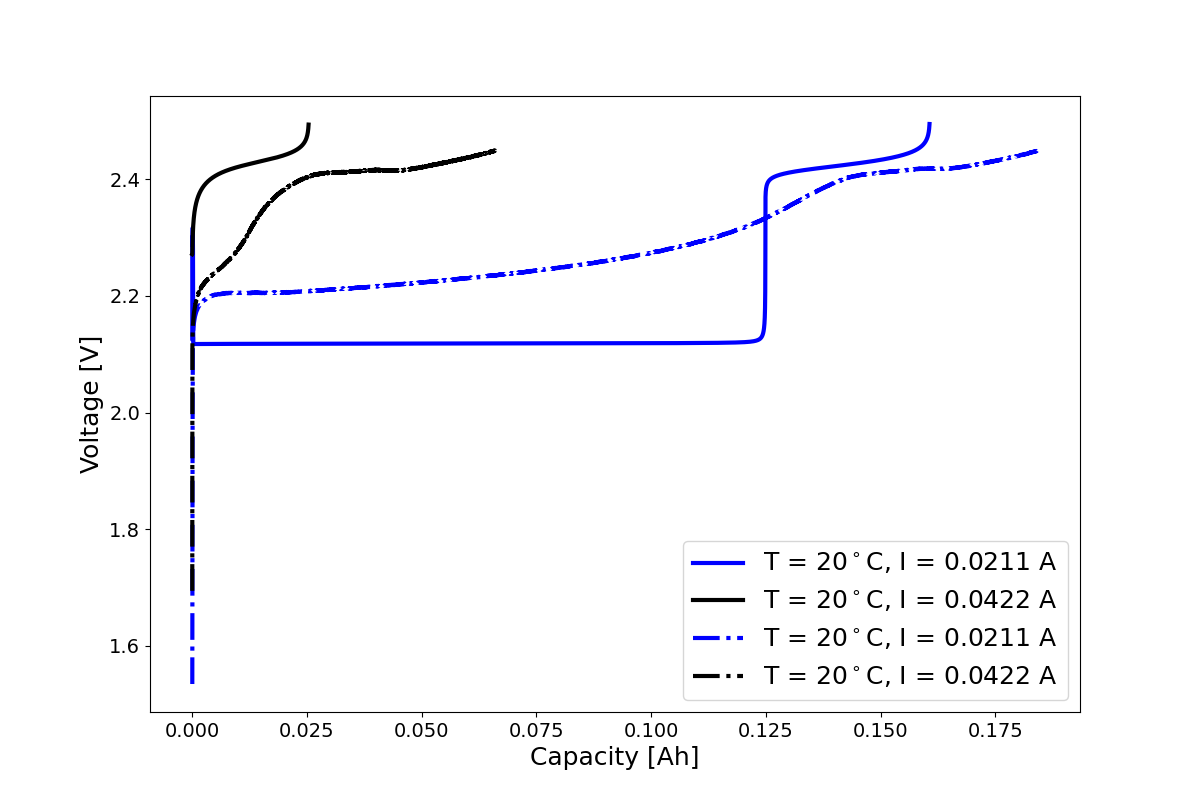}
\caption{Simulated effect charge for 20$^\circ$C cell. Experimental data is represented by dashed lines while solid lines represent the simulated data.}
\label{20_charge}
\end{figure} 
 
\begin{figure}[h]
\includegraphics[width=\columnwidth]{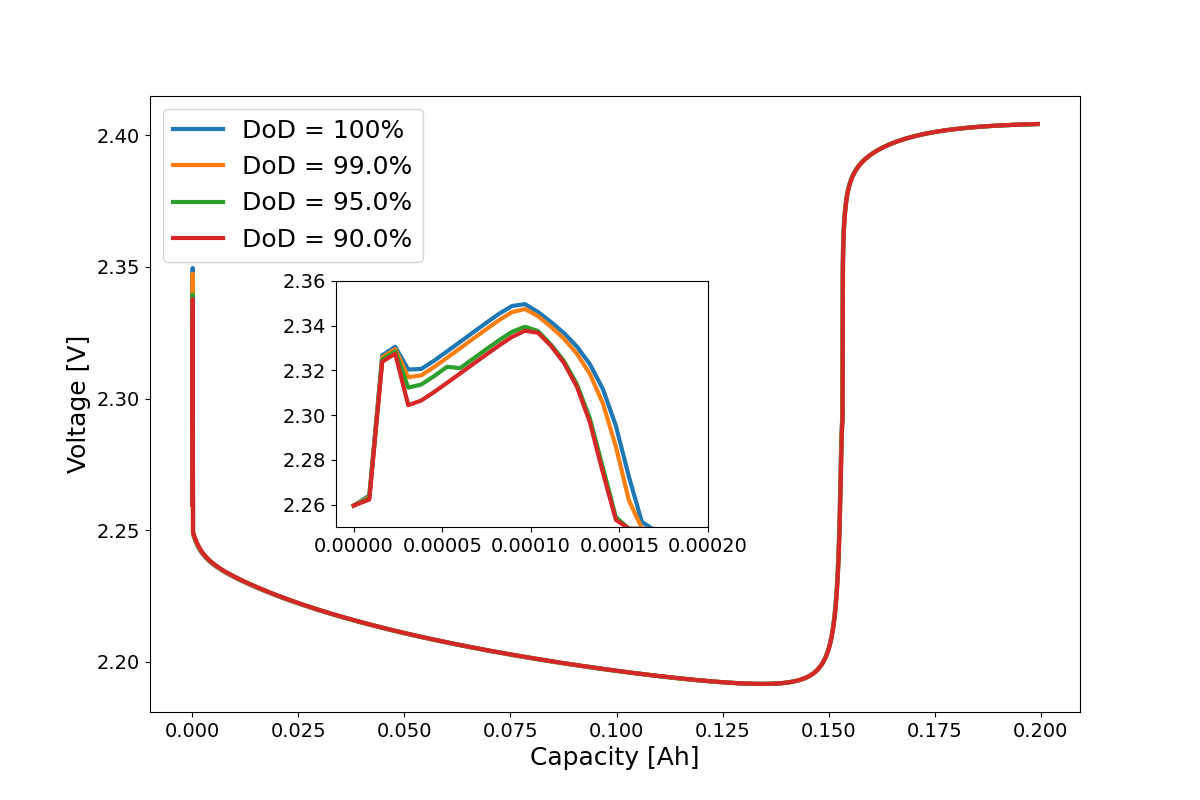}
\caption{Simulated effect of depth-of-discharge (DoD on 40$^\circ$C cell discharged at 0.2C. The voltage spike observed due to DoD can be clearly seen in the inset. The exaggeration of the effect likely due to the sensitivity of the system to low concentrations of Lithium cations at the start of charge along with the simple precipitation model. }
\label{DoD}
\end{figure}

Figure \ref{DoD} displays the effect of DoD on the subsequent initial charge voltage dynamics. The effect on the initial charge kink is very pronounced. This is likely an over-exaggeration of the precipitation feature due to the simple pseudo-spatial model. Nonetheless, this feature was not exhibited in the previous model of Cornish \& Marinescu \cite{cornish2022toward}. Therefore, it is likely that the inclusion of Lithium in the electrochemistry has also helped to make this feature more prominent. On the hand, the precipitation model is quite simple and so a more sophisticated version may yield more reasonable results.

\begin{figure}[h]
\includegraphics[width=\columnwidth]{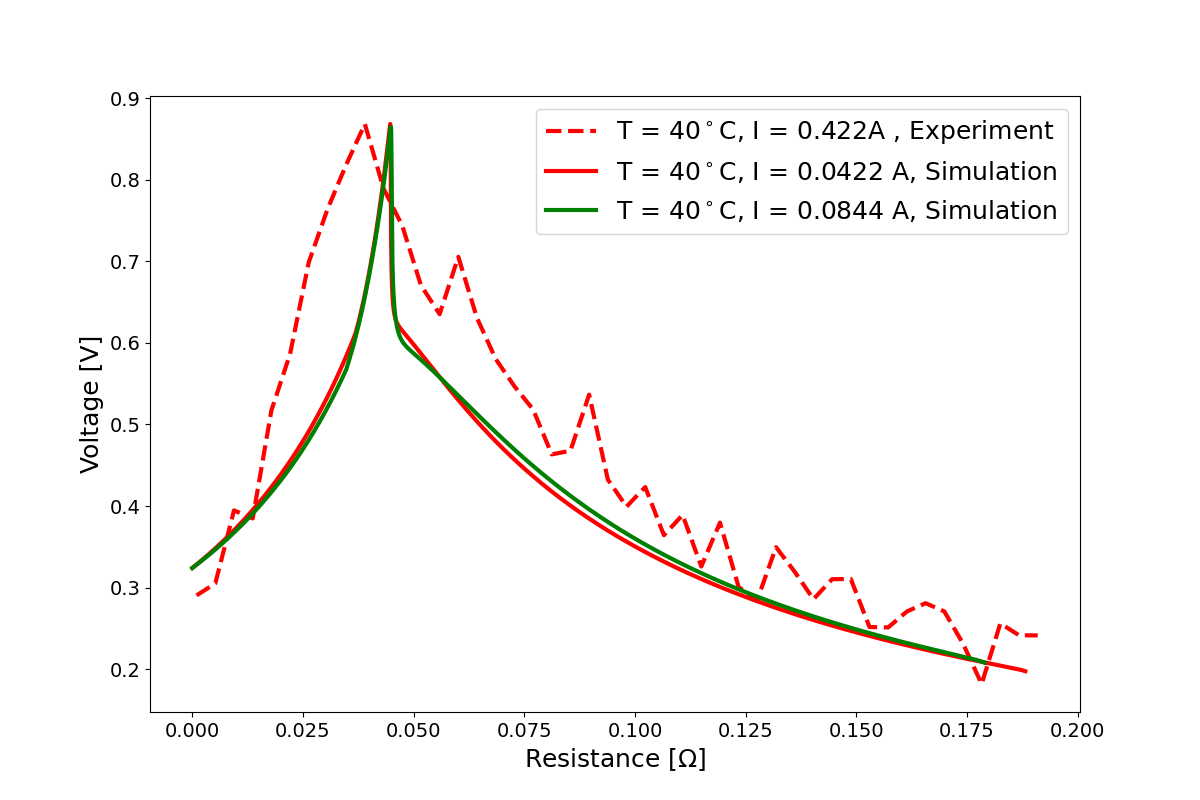}
\caption{Simulated electrolyte resistance for discharge of 40$^\circ$C cell. Experimental data is represented by dashed lines while solid lines represent the simulated data.}
\label{40_resistance}
\end{figure} 

\begin{figure}[h]
\includegraphics[width=\columnwidth]{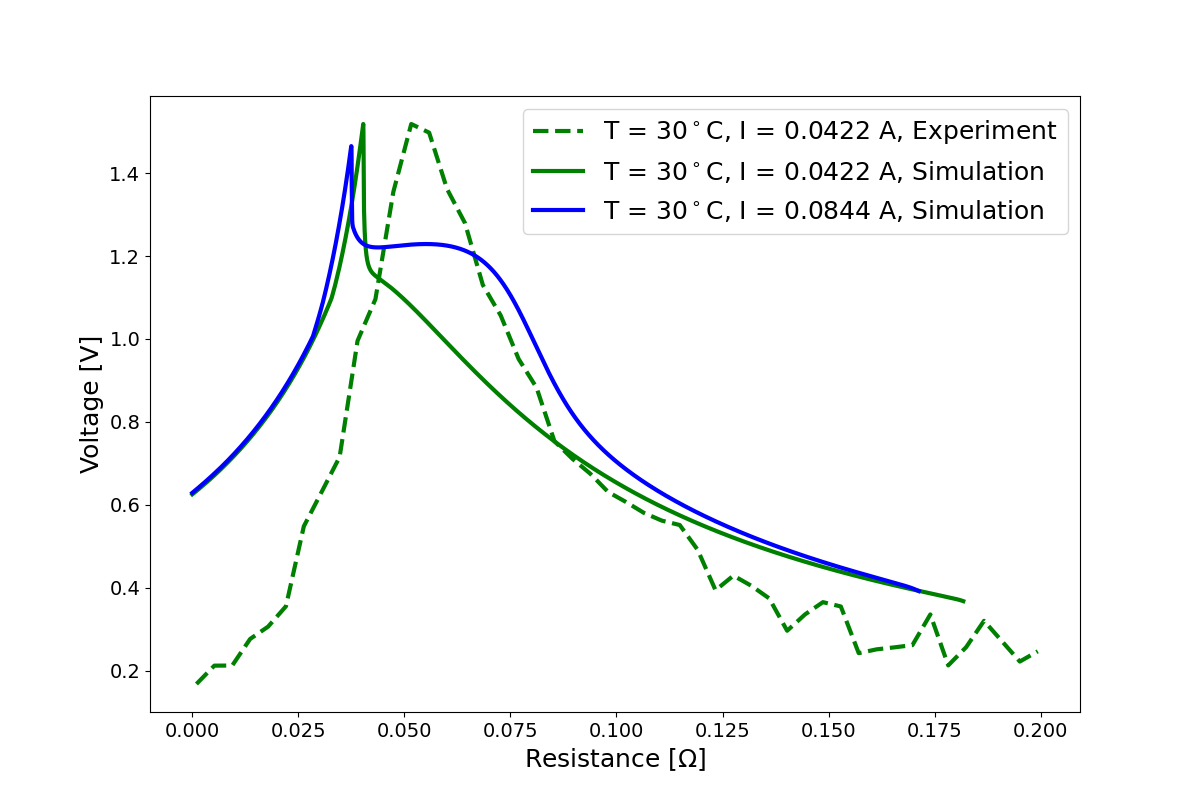}
\caption{Simulated electrolyte resistance for discharge of 30$^\circ$C cell. Experimental data is represented by dashed lines while solid lines represent the simulated data.}
\label{30_resistance}
\end{figure} 

\begin{figure}[h]
\includegraphics[width=\columnwidth]{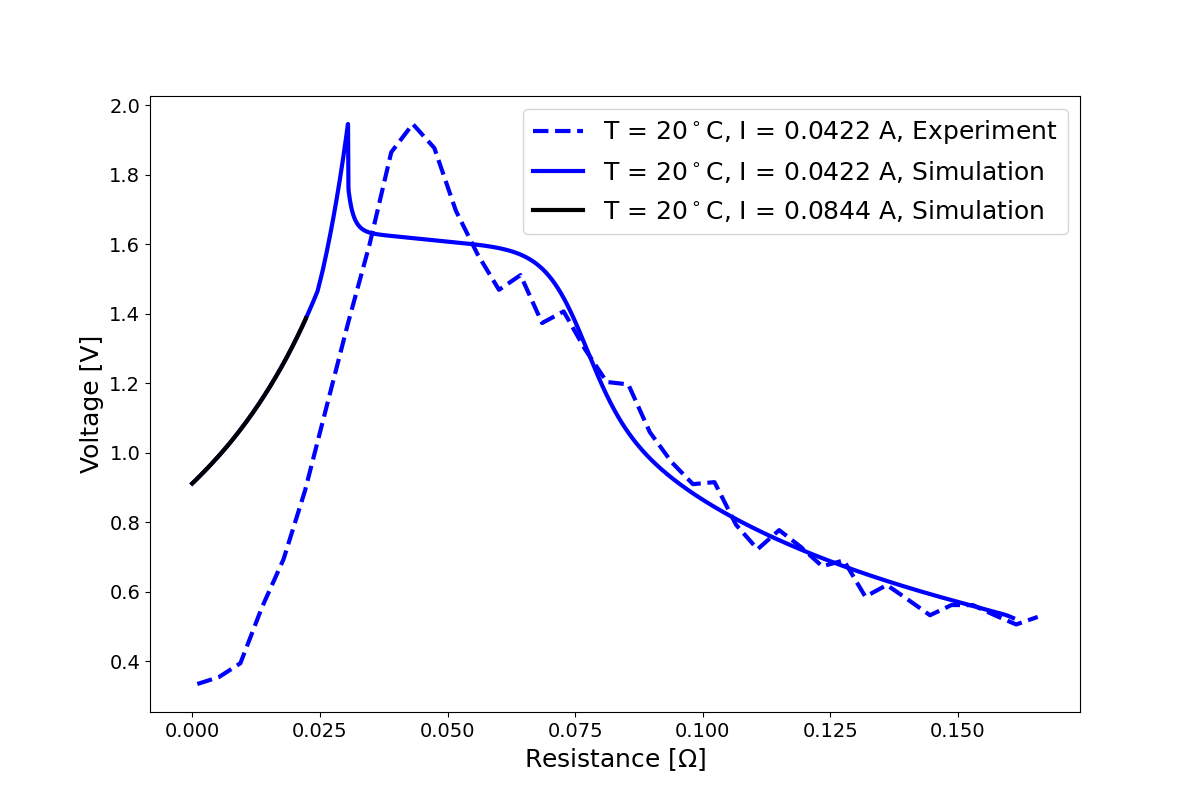}
\caption{Simulated electrolyte resistance for discharge of 20$^\circ$C cell. Experimental data is represented by dashed lines while solid lines represent the simulated data.}
\label{20_resistance}
\end{figure} 

\clearpage
\bibliographystyle{plain}
\bibliography{Modelling_Dynamic_Limitations_of_Lithium_Transport_in_Lithium-Sulfur_Batteries.bib}

\begin{thebibliography}{10}

\bibitem{Andrei2018}
Petru Andrei, Chao Shen, and Jim~P. Zheng.
\newblock {Theoretical and experimental analysis of precipitation and solubility effects in lithium-sulfur batteries}.
\newblock {\em Electrochimica Acta}, 284:469--484, 2018.

\bibitem{Barai2016}
Pallab Barai, Aashutosh Mistry, and Partha~P. Mukherjee.
\newblock {Poromechanical effect in the lithium–sulfur battery cathode}.
\newblock {\em Extreme Mechanics Letters}, 9:359--370, 2016.

\bibitem{boenke2023role}
Tom Boenke, Sebastian Kirchhoff, Florian~S Reuter, Florian Schmidt, Christine Weller, Susanne D{\"o}rfler, Kai Schwedtmann, Paul H{\"a}rtel, Thomas Abendroth, Holger Althues, et~al.
\newblock The role of polysulfide-saturation in electrolytes for high power applications of real world li-s pouch cells.
\newblock {\em Nano Research}, 16(6):8313--8320, 2023.

\bibitem{Bruce2012}
Peter~G. Bruce, Stefan~A. Freunberger, Laurence~J. Hardwick, and Jean~Marie Tarascon.
\newblock {LigO2 and LigS batteries with high energy storage}.
\newblock {\em Nature Materials}, 11(1):19--29, 2012.

\bibitem{cornish2022toward}
Michael Cornish and Monica Marinescu.
\newblock Toward rigorous validation of li-s battery models.
\newblock {\em Journal of The Electrochemical Society}, 169(6):060531, 2022.

\bibitem{Danner2019}
Timo Danner and Arnulf Latz.
\newblock {On the influence of nucleation and growth of S8 and Li2S in lithium-sulfur batteries}.
\newblock {\em Electrochimica Acta}, 322, 2019.

\bibitem{Danner2015}
Timo Danner, Guanchen Zhu, Andreas~F. Hofmann, and Arnulf Latz.
\newblock {Modeling of nano-structured cathodes for improved lithium-sulfur batteries}.
\newblock {\em Electrochimica Acta}, 184:124--133, 2015.

\bibitem{fan2015mechanism}
Frank~Y Fan, W~Craig Carter, and Yet-Ming Chiang.
\newblock Mechanism and kinetics of li2s precipitation in lithium--sulfur batteries.
\newblock {\em Advanced Materials}, 27(35):5203--5209, 2015.

\bibitem{fan2017electrodeposition}
Frank~Y Fan and Yet-Ming Chiang.
\newblock Electrodeposition kinetics in li-s batteries: effects of low electrolyte/sulfur ratios and deposition surface composition.
\newblock {\em Journal of The Electrochemical Society}, 164(4):A917, 2017.

\bibitem{Fan2016}
Frank~Y. Fan, Menghsuan~Sam Pan, Kah~Chun Lau, Rajeev~S. Assary, William~H. Woodford, Larry~A. Curtiss, W.~Craig Carter, and Yet-Ming Chiang.
\newblock {Solvent Effects on Polysulfide Redox Kinetics and Ionic Conductivity in Lithium-Sulfur Batteries}.
\newblock {\em Journal of The Electrochemical Society}, 163(14):A3111--A3116, 2016.

\bibitem{Fronczek2013}
David~N. Fronczek and Wolfgang~G. Bessler.
\newblock {Insight into lithium-sulfur batteries: Elementary kinetic modeling and impedance simulation}.
\newblock {\em Journal of Power Sources}, 244:183--188, 2013.

\bibitem{Ghaznavi2014c}
Mahmoudreza Ghaznavi and P.~Chen.
\newblock {Analysis of a Mathematical Model of Lithium-Sulfur Cells Part III: Electrochemical Reaction Kinetics, Transport Properties and Charging}.
\newblock {\em Electrochimica Acta}, 137:575--585, 2014.

\bibitem{Ghaznavi2014b}
Mahmoudreza Ghaznavi and P.~Chen.
\newblock {Sensitivity analysis of a mathematical model of lithium-sulfur cells: Part II: Precipitation reaction kinetics and sulfur content}.
\newblock {\em Journal of Power Sources}, 257:402--411, 2014.

\bibitem{gupta2020influence}
Abhay Gupta, Amruth Bhargav, John-Paul Jones, Ratnakumar~V Bugga, and Arumugam Manthiram.
\newblock Influence of lithium polysulfide clustering on the kinetics of electrochemical conversion in lithium--sulfur batteries.
\newblock {\em Chemistry of Materials}, 32(5):2070--2077, 2020.

\bibitem{Hofmann2014}
Andreas~F. Hofmann, David~N. Fronczek, and Wolfgang~G. Bessler.
\newblock {Mechanistic modeling of polysulfide shuttle and capacity loss in lithium-sulfur batteries}.
\newblock {\em Journal of Power Sources}, 259:300--310, 2014.

\bibitem{Hua2019}
Xiao Hua, Teng Zhang, Gregory~J. Offer, and Monica Marinescu.
\newblock {Towards online tracking of the shuttle effect in lithium sulfur batteries using differential thermal voltammetry}.
\newblock {\em Journal of Energy Storage}, 21(January):765--772, 2019.

\bibitem{Hunt2018}
I.~Hunt, T.~Zhang, Y.~Patel, M.~Marinescu, R.~Purkayastha, P.~Kovacik, S.~Walus, A.~Swiatek, and G.~J. Offer.
\newblock {The Effect of Current Inhomogeneity on the Performance and Degradation of Li-S Batteries}.
\newblock {\em Journal of The Electrochemical Society}, 165(1):A6073--A6080, 2018.

\bibitem{kirchhoff2023evaluation}
Sebastian Kirchhoff, Christian Leibing, Paul H{\"a}rtel, Thomas Abendroth, Susanne D{\"o}rfler, Holger Althues, Stefan Kaskel, and Andrea Balducci.
\newblock Evaluation of glyoxal-based electrolytes for lithium-sulfur batteries.
\newblock {\em Batteries}, 9(4):210, 2023.

\bibitem{Kumaresan2008}
Karthikeyan Kumaresan, Yuriy Mikhaylik, and Ralph~E. White.
\newblock {A mathematical model for a lithium-sulfur cell}.
\newblock {\em Journal of the Electrochemical Society}, 155(8), 2008.

\bibitem{Marinescu2018}
Monica Marinescu, Laura O'Neill, Teng Zhang, Sylwia Walus, Timothy~E. Wilson, and Gregory~J. Offer.
\newblock {Irreversible vs Reversible Capacity Fade of Lithium-Sulfur Batteries during Cycling: The Effects of Precipitation and Shuttle}.
\newblock {\em Journal of The Electrochemical Society}, 165(1):A6107--A6118, 2018.

\bibitem{Marinescu2016}
Monica Marinescu, Teng Zhang, and Gregory~J. Offer.
\newblock {A zero dimensional model of lithium-sulfur batteries during charge and discharge}.
\newblock {\em Physical Chemistry Chemical Physics}, 18(1):584--593, 2016.

\bibitem{Mikhaylik2004}
Yuriy~V. Mikhaylik and James~R. Akridge.
\newblock {Polysulfide Shuttle Study in the Li/S Battery System}.
\newblock {\em Journal of The Electrochemical Society}, 151(11):A1969, 2004.

\bibitem{Neidhardt2012}
Jonathan~P. Neidhardt, David~N. Fronczek, Thomas Jahnke, Timo Danner, Birger Horstmann, and Wolfgang~G. Bessler.
\newblock {A Flexible Framework for Modeling Multiple Solid, Liquid and Gaseous Phases in Batteries and Fuel Cells}.
\newblock {\em Journal of The Electrochemical Society}, 159(9):A1528--A1542, 2012.

\bibitem{Parke2020}
Caitlin~D. Parke, Akshay Subramaniam, Suryanarayana Kolluri, Daniel~T. Schwartz, and Venkat~R. Subramanian.
\newblock {An Efficient Electrochemical Tanks-in-Series Model for Lithium Sulfur Batteries}.
\newblock {\em Journal of The Electrochemical Society}, 167(16):163503, 2020.

\bibitem{Parke2021}
Caitlin~D. Parke, Akshay Subramaniam, Venkat~R. Subramanian, and Daniel~T. Schwartz.
\newblock {Realigning the Chemistry and Parameterization of Lithium-Sulfur Battery Models to Accommodate Emerging Experimental Evidence and Cell Configurations}.
\newblock {\em ChemElectroChem}, 8(6):1098--1106, 2021.

\bibitem{Ren2016}
Y.~X. Ren, T.~S. Zhao, M.~Liu, P.~Tan, and Y.~K. Zeng.
\newblock {Modeling of lithium-sulfur batteries incorporating the effect of Li2S precipitation}.
\newblock {\em Journal of Power Sources}, 336:115--125, 2016.

\bibitem{Song2013}
Min~Kyu Song, Elton~J. Cairns, and Yuegang Zhang.
\newblock {Lithium/sulfur batteries with high specific energy: Old challenges and new opportunities}.
\newblock {\em Nanoscale}, 5(6):2186--2204, 2013.

\bibitem{Thangavel2016}
Vigneshwaran Thangavel, Kan-Hao Xue, Youcef Mammeri, Matias Quiroga, Afef Mastouri, Claude Gu{\'{e}}ry, Patrik Johansson, Mathieu Morcrette, and Alejandro~A. Franco.
\newblock {A Microstructurally Resolved Model for Li-S Batteries Assessing the Impact of the Cathode Design on the Discharge Performance}.
\newblock {\em Journal of The Electrochemical Society}, 2016.

\bibitem{Zhang2015}
Teng Zhang, Monica Marinescu, Laura O'Neill, Mark Wild, and Gregory Offer.
\newblock {Modeling the voltage loss mechanisms in lithium-sulfur cells: The importance of electrolyte resistance and precipitation kinetics}.
\newblock {\em Physical Chemistry Chemical Physics}, 17(35):22581--22586, 2015.

\bibitem{Zhang2016}
Teng Zhang, Monica Marinescu, Sylwia Walus, and Gregory~J. Offer.
\newblock {Modelling transport-limited discharge capacity of lithium-sulfur cells}.
\newblock {\em Electrochimica Acta}, 219:502--508, 2016.

\end{thebibliography}

\end{document}